\documentclass[aps,reprint,floatfix,superscriptaddress]{revtex4-2}
\usepackage{dcolumn}
\usepackage[colorlinks=true, citecolor=blue, linkcolor=blue, urlcolor=blue]{hyperref}
\usepackage{comment}
\usepackage{tikz}
\usepackage{amsmath}
\usepackage{makecell}
\usepackage{tabularx}
\usepackage{orcidlink}
\AtBeginDocument{}

\definecolor{gold}{rgb}{1.0, 0.84, 0.0}
\definecolor{skyblue}{RGB}{135,206,235}

\begin{document}

\title{Charge Transport and Multiplication in Lateral Amorphous Selenium Devices Under
Cryogenic Conditions}

\author{M.~Rooks\,\orcidlink{0000-0002-1629-6825}}
\email{Corresponding Author msrooks@gmail.com}
\affiliation{Department of Physics, University of Texas at Arlington, Arlington, TX 76019, USA}
\affiliation{Physics Division, Oak Ridge National Laboratory, Oak Ridge, TN 37831, USA}

\author{S.~Abbaszadeh}
\affiliation{Department of Electrical and Computer Engineering, University of California, Santa Cruz, CA 95064, USA}

\author{J.~Asaadi}
\affiliation{Department of Physics, University of Texas at Arlington, Arlington, TX 76019, USA}

\author{V.~A.~Chirayath}
\affiliation{Department of Physics, University of Texas at Arlington, Arlington, TX 76019, USA}

\author{M.~Á.~García-Peris}
\affiliation{Department of Physics and Astronomy, University of Manchester, Manchester M13 9PL, United Kingdom}

\author{E.~Gramellini}
\affiliation{Department of Physics and Astronomy, University of Manchester, Manchester M13 9PL, United Kingdom}

\author{K.~Hellier}
\affiliation{Department of Electrical and Computer Engineering, University of California, Santa Cruz, CA 95064, USA}

\author{B.~Sudarsan\,\orcidlink{0000-0002-6228-5043}}
\affiliation{Department of Physics, Florida State University, Tallahassee, FL 32306, USA}

\author{I.~Tzoka\,\orcidlink{0000-0001-7811-5068}}
\affiliation{Department of Physics, University of Texas at Arlington, Arlington, TX 76019, USA}

\date{\today}

\begin{abstract}
Cryogenic photon sensing for high-energy physics motivates photosensor technologies that combine large-area scalability with internal gain and stable operation at low temperature. Amorphous selenium is a promising photoconductor, yet its field- and temperature-dependent transport and avalanche response in lateral geometries have not been systematically established. This work reports field-resolved photocurrent measurements of lateral a-Se devices from 93 K to 297 K under 401 nm excitation at fields up to 120 V/\textmu m. Below avalanche onset, the external quantum efficiency was described by the Onsager model, yielding effective post-thermalization separations that decrease with decreasing temperature. The field-assisted detrapping region was evaluated using several transport models, with the data favoring field-assisted hopping and thermally-assisted tunneling as the mechanisms that best capture the temperature evolution of the photocurrent. The boundaries between field-assisted detrapping, transport-limited conduction, and avalanche shift with temperature; at 93 K the response transitions directly from detrapping into avalanche. Avalanche multiplication was analyzed using the Lucky-drift model. These results provide the first systematic characterization of cryogenic avalanche behavior in lateral a-Se detectors and establish quantitative trends relevant to low-temperature, high-gain photodetector design.
\end{abstract}

\maketitle
\section{Introduction}\label{sec:Intro}

Photon detection plays a central role in applications ranging from high-energy physics (HEP) to medical imaging and environmental monitoring. Conventional photosensors for HEP often employ photomultiplier tubes or solid-state devices such as silicon photomultipliers, both of which are well established in nuclear and particle physics as well as in medical and radiation detection systems~\cite{Ota2021}. Semiconductor materials such as silicon, cadmium telluride, and cadmium zinc telluride offer high carrier mobility and efficient charge collection, but their scalability and fabrication costs can present challenges for large-area or cost-sensitive deployments~\cite{Schlomka2008,Bornefalk2010,Kuvvetli2010}.

Amorphous selenium (a-Se) provides an alternative semiconductor detection medium that enables scintillator-free, direct photon-to-charge conversion in a photoconductor, can support high spatial resolution, and is compatible with large-area, low-cost manufacturing~\cite{Stavro2018}. While its most mature applications are in medical x-ray imaging, a-Se has also been developed for high-gain avalanche photodetectors and other emerging photon detection systems~\cite{stavro2016,sensors2013,hellier2023}. Demonstrations of stable operation in cryogenic environments and sensitivity extending into the vacuum ultraviolet have further expanded its potential for use in noble-liquid detectors~\cite{Rooks2023,Rooks2026}.

A detailed understanding of a-Se photodetector behavior across applied field and temperature is essential to support these applications. In particular, the coupled field- and temperature-dependent behavior of photogeneration efficiency, carrier transport, and avalanche multiplication remains incompletely characterized in lateral geometries under pulsed optical excitation at cryogenic temperatures and high fields. Prior studies of lateral a-Se devices have primarily focused on steady-state photoresponse, dark-current suppression, and geometric optimization, with less emphasis on avalanche behavior~\cite{Chang2016,Abbaszadeh2013_2,Wang2010}. Avalanche behavior in laterally structured a-Se detectors at cryogenic temperature was initially reported in an instrumentation study~\cite{Rooks2026}. The present work extends that result by incorporating additional data from the same devices and providing a systematic investigation of charge transport and avalanche multiplication as functions of temperature and applied field in lateral devices.

Photoconductive response was investigated for lateral a-Se detectors fabricated on interdigitated electrode (IDE) structures incorporating a thin polyimide (PI) blocking layer. The study focuses on stabilized a-Se devices operated over a temperature range of 93--297\,K and applied fields of 10--120\,V/\textmu m. Pulsed 401\,nm laser illumination was used to generate carriers, and the resulting signals were read out using a charge-sensitive preamplifier. These measurements enable quantitative analysis of photogeneration efficiency, charge transport, and avalanche multiplication over a wide range of fields and temperatures. Photogeneration was modeled using the Onsager formalism for field-assisted geminate-pair dissociation, charge transport was analyzed to identify temperature-dependent transitions in the dominant conduction mechanisms, and avalanche multiplication was characterized using the Lucky-drift (LD) model to extract impact-ionization coefficients~\cite{Kasap2004}.

\section{Device Design}\label{sec:device}

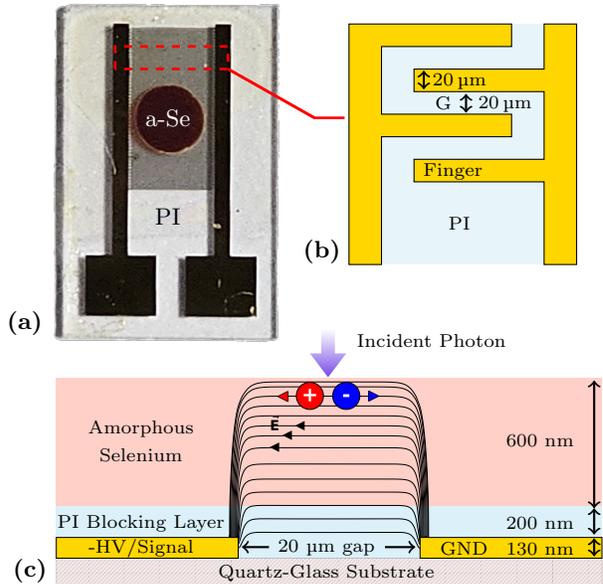
\begin{figure}
\centering
\makebox[\columnwidth][l]{%
    \hspace{5mm} 
    \begin{minipage}[t][6cm][c]{0.36\columnwidth}
        \begin{tikzpicture}[remember picture]
            \node[anchor=south west, inner sep=0] (img) at (0,0)
                {\includegraphics[width=\linewidth]{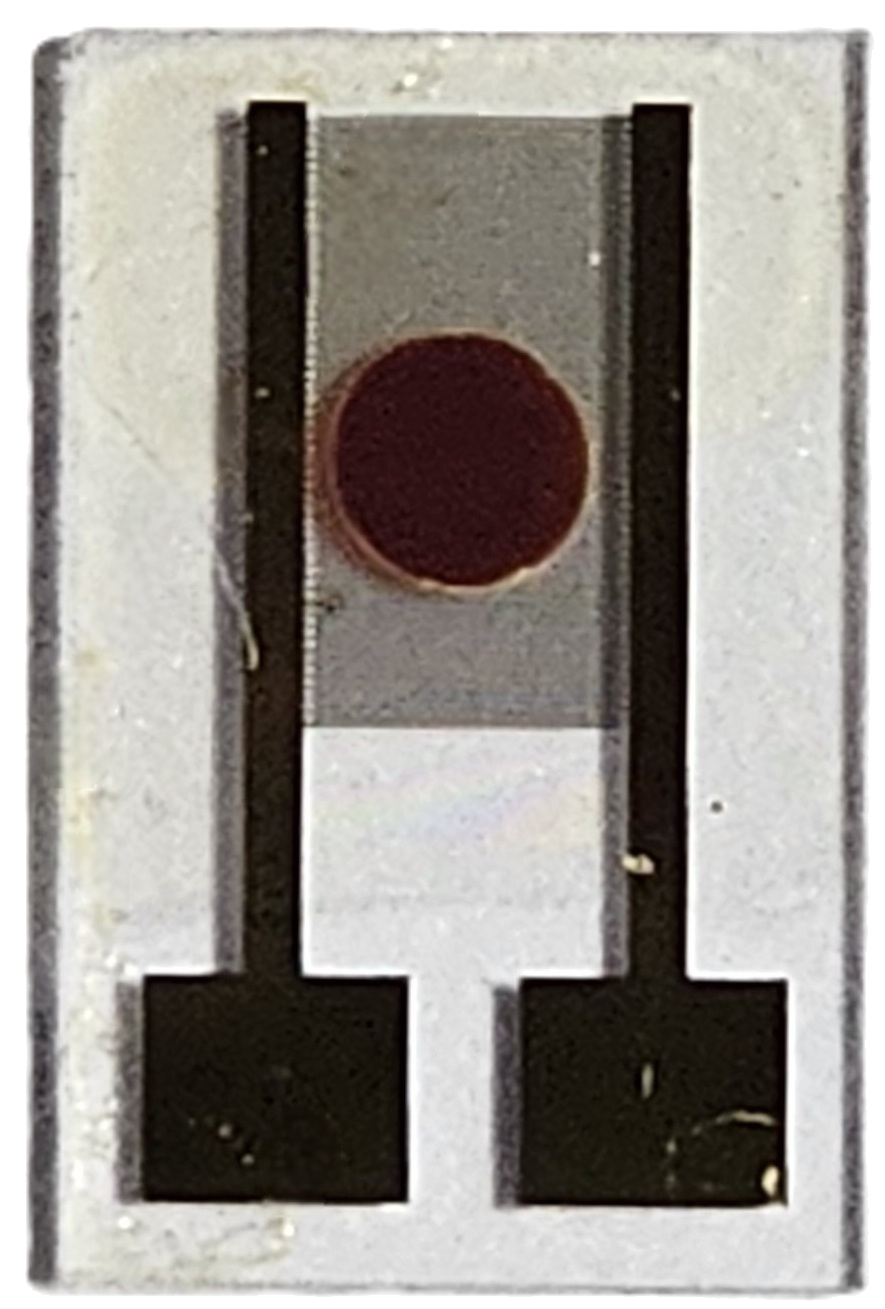}};
            \node[text=white] at (1.6,3.01) {a-Se}; 
            \node[text=black] at (1.6,1.7) {PI};
            \draw[red, dashed, line width=1pt] (0.9,3.66) rectangle ++(1.5,0.3);
            \coordinate (rectBR) at (2.4,3.66);

            \node[anchor=north east,font=\bfseries,overlay] 
              at ([xshift=0.5mm,yshift=5.5mm]img.south west) {(a)};
        \end{tikzpicture}
    \end{minipage}%
    \hspace{5mm}
    \begin{minipage}[t][6cm][c]{0.60\columnwidth}
        \begin{tikzpicture}[x=\linewidth/6, y=6cm/6,remember picture]
            
            \path[use as bounding box] (0,0) rectangle (5.9,4.5);

            \fill[skyblue, fill opacity=0.2]
              (0.9,1) -- (3.4,1)  -- (3.4,4.2) -- (0.9,4.2)  -- cycle;

            \filldraw[black, thin, fill=gold] 
              (0.4,1) -- (0.9,1) -- (0.9,2.7) -- (2.9,2.7) -- (2.9,3.0) -- (0.9,3.0)
              -- (0.9,3.9) -- (2.9,3.9) -- (2.9,4.2) -- (0.9,4.2) -- (0.4,4.2) -- cycle;
            
            \draw[black, thin, fill=gold] 
              (3.4,1) -- (3.9,1) -- (3.9,4.2) -- (3.4,4.2)  
              -- (3.4,3.6) -- (1.4,3.6) -- (1.4,3.3) -- (3.4,3.3)  
              -- (3.4,2.4) -- (1.4,2.4) -- (1.4,2.1) -- (3.4,2.1)  
              -- cycle;
            
            \draw[<->, thick] (1.55,3.33) -- (1.55,3.58);
            \node at (2.1,3.44) {\scriptsize 20\,\textmu m};

            \node at (2.8,3.13) {\scriptsize 20\,\textmu m};
            \draw[<->, thick] (2.2,3.02) -- (2.2,3.27);
            \node at (1.85,3.15) {\scriptsize G};
            \node at (2,2.24) {\scriptsize Finger};
            \node at (2.1,1.55) {\scriptsize PI};

            \node[anchor=north east,font=\bfseries,overlay] 
              at ([xshift=3.5mm,yshift=14.5mm]current bounding box.south west) {(b)};
            
        \end{tikzpicture}
    \end{minipage}%

    \begin{tikzpicture}[remember picture, overlay]
      \draw[red, line width=1pt] (rectBR) -- ++(1.15,-0.65);
      \draw[red, line width=1pt] (rectBR) -- ++(1.15,-0.65) -- ++(0.4,0);
    \end{tikzpicture}
}

\vspace{-1cm}

\begin{tikzpicture}[x=\columnwidth/6, y=4cm/6,remember picture]
    \node[anchor=south west, inner sep=0] (ide) at (0,0)
        {\includegraphics[width=0.85\columnwidth]{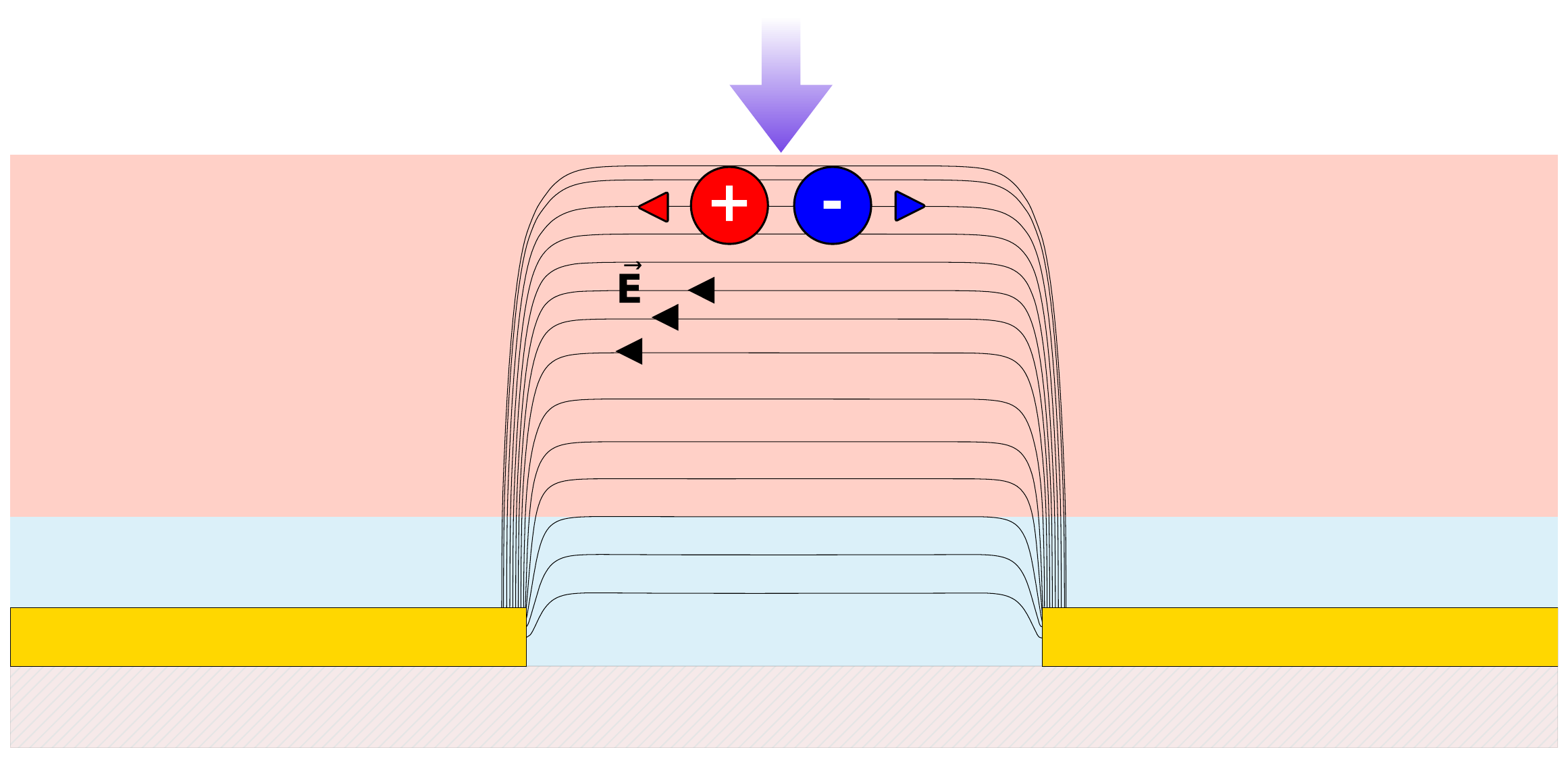}};
    \node[text=black] at (0.82,1.35) {\scriptsize PI Blocking Layer};
    \node[text=black] at (4.5,1.35) {\scriptsize 200 nm};
    \draw[<->, thick] (5,1.125) -- (5,1.65);

    \node[text=black] at (0.82,0.85) {\scriptsize -HV/Signal};
    \node[text=black] at (3.8,0.85) {\scriptsize GND};
    \node[text=black] at (4.5,0.85) {\scriptsize 130 nm};
    \draw[<->, thick] (5,0.69) -- (5,1.05);    

    \node[text=black, align=center] at (0.82,3) {\scriptsize Amorphous \\ \scriptsize Selenium};
    \node[text=black] at (4.5,3) {\scriptsize 600 nm};
    \draw[<->, thick] (5,1.65) -- (5,4.2);    

    \node[text=black] at (2.53,0.35) {\scriptsize Quartz-Glass Substrate};
    \node[text=black] at (3.5, 5) {\scriptsize Incident Photon};

    \node[text=black] at (2.53,0.875) {\scriptsize 20 \textmu m gap};    
    \draw[<-, thick] (1.75,0.875) -- (2,0.875); 
    \draw[->, thick] (3.07,0.875) -- (3.35,0.875);
    \node[anchor=north east,font=\bfseries,overlay] 
      at ([xshift=0.5mm,yshift=5.5mm]ide.south west) {(c)};
\end{tikzpicture}

\caption{(a) Fabricated detector showing a 1.6\,mm diameter, 600\,nm-thick a-Se dot thermally evaporated onto a 200\,nm-thick spin-coated PI layer covering the entire IDE region. (b) Schematic of the IDE geometry highlighting the electrode finger width and gap separation, G. (c) Simulated electric field distribution in the a-Se photodetector at a nominal applied field of 100\,V/\textmu m. Field lines are shown only within the active a-Se region for clarity. Adapted from~\cite{Rooks2026}.}
\label{fig:IDE}
\end{figure}

The detectors used in this study were fabricated on IDE structures with 20\,\textmu m finger width and a 20\,\textmu m gap as shown in Fig.~\ref{fig:IDE}(b). A 200\,nm-thick polyimide blocking layer was spin-coated over the entire IDE region to suppress charge injection from the electrodes. Stabilized a-Se was then thermally evaporated through a shadow mask to form a 600\,nm-thick dot centered on the electrode array. The dot was 1.6\,mm in diameter and defined the optically sensitive area of the device. Under applied bias, photogenerated holes drift laterally within the a-Se film, parallel to the electrode gap, toward the signal electrode. Holes dominate transport in a-Se. The hole drift mobility is \(\sim 0.13\text{--}0.14\,\mathrm{cm^2\,V^{-1}\,s^{-1}}\), which greatly exceeds the electron drift mobility of \(\sim 0.005\text{--}0.007\,\mathrm{cm^2\,V^{-1}\,s^{-1}}\)~\cite{kasap2009}. Figures~\ref{fig:IDE}(a) and~\ref{fig:IDE}(b) show an image of the fabricated detector and a schematic of the IDE layout, respectively. The simulated electric-field distribution for a nominal applied field of 100\,V/\textmu m, computed using COMSOL Multiphysics\textsuperscript{\textregistered}, is shown in Fig.~\ref{fig:IDE}(c). Full fabrication details and measured as-fabricated dimensions are reported in~\cite{Rooks2026}. The dimensions quoted here and in Fig.~\ref{fig:IDE} are nominal layout targets; the measured values differ from nominal by less than 3\%, and only details relevant to device operation and modeling are summarized here.

The PI layer insulated both IDE electrodes, preventing direct carrier transfer into the metal. It therefore acts as a blocking boundary at both electrodes, suppressing injection and limiting hole extraction at the collecting side. As a result, the readout is capacitively coupled through the PI, and the observed waveform is dominated by Ramo–Shockley induced current rather than steady-state conduction.

\section{Experimental Setup}
\label{sec:methods}

\begin{figure}
\centering
\includegraphics[width=\columnwidth, trim=3.8cm 0 0 0, clip]{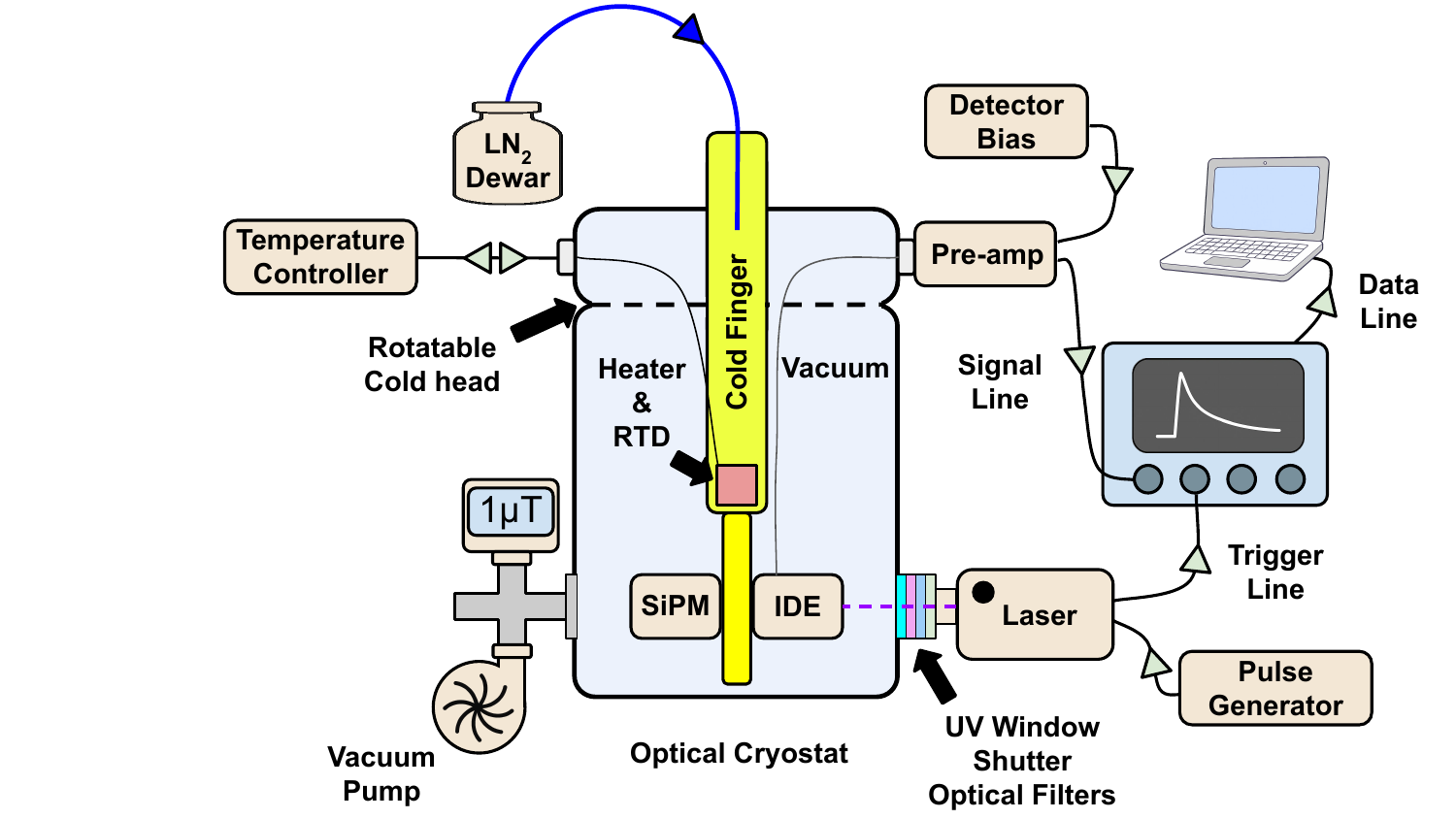}
\caption{Diagram of the optical cryostat test setup. A picosecond-pulsed diode laser at 401\,nm is directed through a fused-silica window to illuminate the a-Se detector mounted on a copper block at the cold finger. The detector is biased through external feedthroughs, and the output is routed to a charge-sensitive preamplifier and oscilloscope for signal acquisition. Adapted from~\cite{Rooks2026}.}
\label{fig:expSetup}
\end{figure}

Measurements were performed in a liquid-nitrogen-cooled optical cryostat operated under vacuum. Each photodetector was mounted to a cold finger using a custom connector that positioned the detector on a copper block to aid thermal transfer. The devices were operated at discrete temperatures of 93\,K, 165\,K, 200\,K, and 297\,K, with temperature stability verified prior to data acquisition. A schematic of the optical cryostat test setup is shown in Fig.~\ref{fig:expSetup}. The experimental setup and calibration procedures are described in detail in~\cite{Rooks2026}.

Excitation was provided by a 401\,nm picosecond-pulsed diode laser operating at 4\,Hz. The beam was collimated to a diameter of 1.2\,mm, directed through a fused-silica window and aligned to the center of the a-Se dot. The incident flux was fixed at 1.56\,M photons (0.74\,pJ) per pulse with a systematic uncertainty of 5\%~\cite{Rooks2026}. Applied fields in the range of 10--120\,V/\textmu m were established by biasing the photodetector with an external voltage source. Detector response was processed with a charge-sensitive preamplifier and recorded on a digital oscilloscope. Waveforms were stored for offline analysis.

\section{Theoretical background}
\label{sec:Theory}

The theoretical background is organized as follows. External quantum efficiency (EQE) is defined first in Sec.~\ref{sec:eqe_theory}. Geminate pair dissociation in a-Se is introduced next in Sec.~\ref{sec:geminate} as a low-field description of carrier escape. Avalanche multiplication at high fields is described using the LD model in Sec.~\ref{sec:AVLD}. The section concludes with charge transport in a-Se in Sec.~\ref{sec:transport}, describing the evolution of the response with increasing field through the pre-avalanche transport interval, beginning with the transport formalism and effective mobility, and then outlining trap-limited drift in extended states, hopping transport between localized states, and field-enhanced emission involving tunneling.

\subsection{External quantum efficiency}
\label{sec:eqe_theory}

The EQE is defined as the ratio of collected charge carriers to incident photons at the detector surface. In a-Se--based devices, it can be expressed as
\begin{equation}
\mathrm{EQE}(E,T)=\eta(E,T)\,\xi(E,T),
\label{eq:eqe_decomp}
\end{equation}

where \(\eta(E,T)\) is the field- and temperature-dependent photogeneration efficiency per absorbed photon, and \(\xi(E,T)\) is the charge-collection efficiency, defined as the fraction of photogenerated carriers that contribute to the measured response.

The photogeneration efficiency \(\eta(E,T)\) is determined by the dissociation probability of geminate electron--hole pairs. The charge collection efficiency \(\xi(E,T)\) includes losses due to trapping, recombination, and device geometry, and accounts for the overall drift transport across the photoconductor. This decomposition highlights that EQE is not solely set by intrinsic absorption but reflects the coupled field and temperature dependences of photogeneration and charge collection.

\subsection{Geminate pair dissociation in a-Se}
\label{sec:geminate}

Photogeneration in a-Se begins with the creation of thermalized electron--hole pairs bound by Coulomb attraction. Whether these geminate pairs dissociate into free carriers depends on their initial separation, the applied electric field, and the thermal energy of the system. The dissociation process is described by the Onsager model, which gives the probability that a pair escapes recombination via field-assisted diffusion~\cite{Onsager1938,Pai1975}.

The field- and temperature-dependent photogeneration efficiency can be evaluated using the integral form of the Onsager model given by~\cite{yip1981}:
\begin{equation}
\eta = \eta_0\!\left[1 - \frac{1}{2}\int_0^2 e^{-\gamma y}\,dy
\int_0^D I_0(2\sqrt{\gamma y x})\,e^{-x}\,dx \right]
\label{eq:yip1}
\end{equation}
with
\begin{equation}
\gamma = \frac{q E r_0}{2 k_B T},\; D = \frac{r_c}{r_0},\;
r_c = \frac{q^2}{4 \pi \varepsilon_r \varepsilon_0 k_B T}.
\label{eq:yip2}
\end{equation}

Here \(\eta_0\) is a material-dependent, field-independent efficiency representing the fraction of absorbed photons that produce thermalized electron--hole pairs. The parameter \(r_0\) is the initial pair separation, \(I_0\) is the zeroth-order modified Bessel function of the first kind, q is the elementary charge, \(E\) is the electric field, \(k_B\) is Boltzmann’s constant, \(T\) is the temperature, \(\varepsilon_r\) is the relative permittivity of the medium, and \(\varepsilon_0\) is the permittivity of free space. For stabilized a-Se, \(\varepsilon_r\) is reported as 6.7 \cite{Abbaszadeh2013_2,Frey}.

This expression captures the dependence of geminate pair dissociation on both field and temperature. Through the field dependence of \(\gamma\) and the temperature dependence of \(r_c\), the Onsager model predicts that the escape probability and thus \(\eta\) increase with field and decrease at lower temperature. Because EQE as defined in Eq.~\ref{eq:eqe_decomp} includes the collection efficiency, the observed EQE additionally depends on \(\xi(E,T)\), and is therefore not determined by the Onsager model alone.

\subsection{Avalanche and the Lucky-drift Model}
\label{sec:AVLD}

At sufficiently high electric fields, typically above 70\,V/\textmu m at room temperature, photogenerated carriers in a-Se undergo impact ionization, giving rise to avalanche multiplication~\cite{Kasap2004,Tanaka2014}. In a conventional vertical photoconductor, the a-Se film thickness \(d\) represents the total drift distance and the multiplication factor \(M\) is related to \(d\) by
\begin{equation}
M = e^{\alpha(E)d},
\label{eq:Malpha}
\end{equation}
where \(\alpha(E)\) is the field-dependent ionization coefficient~\cite{Kasap2004}. 

In lateral device geometries, carrier generation is distributed across the electrode gap 
\(G\) shown in Fig.~\ref{fig:IDE}, and the multiplication must account for variable ionization probability over this distance. Assuming a uniform electric field across the gap, a carrier generated at position \(x\) must drift the remaining distance \((G-x)\) to the collecting electrode, giving a position-dependent multiplication factor
\begin{equation}
M(x) = e^{\alpha(E)(G-x)}.
\label{eq:Mx}
\end{equation}
Averaging over a uniform distribution of generation positions within \([0,G]\) yields
\begin{equation}
\langle M \rangle = \frac{1}{G}\int_{0}^{G} e^{\alpha(E)(G-x)}\,dx
   = \frac{e^{\alpha(E)G}-1}{\alpha(E)G}.
\label{eq:Mavg}
\end{equation}

As shown in Fig.~\ref{fig:IDE}(c), simulations indicate that the field in the near-surface a-Se region relevant for 401\,nm absorption is approximately uniform across the central portion of the electrode gap. Field reductions are confined to narrow regions near the electrode edges. Near the hole-collecting \(-\)HV/Signal side, the field recovers to within 5\% of the central-gap value by 0.57\,\textmu m, with a mean reduction of 13.3\% relative to the central-gap field, limiting its impact on the gap-averaged multiplication.

The field dependence of $\alpha(E)$ in disordered semiconductors is described by the LD model, in which carriers accelerate between energy-relaxing collisions and their trajectories are randomized by momentum-relaxing scattering~\cite{Kasap2004}. Carriers gain energy from the applied field between successive energy-relaxing events and lose energy to the lattice through phonon scattering in the amorphous network~\cite{Kasap2004,Rubel2004,Tanaka2014}. In the LD formulation, scattering is treated in two effective classes: momentum-relaxing events that randomize the trajectory without significant energy loss, and rarer energy-relaxing events that thermalize the carrier energy.

The lucky ballistic probability, following Shockley’s lucky electron argument, describes the likelihood that a carrier attains energy $\mathcal{E}$ without a momentum-relaxing collision~\cite{Shockley1961,Ridley1983}:
\begin{equation}
P_{\mathrm{LB}}(\mathcal{E})=
\exp\!\left[-\int_{0}^{\mathcal{E}}\frac{d\mathcal{E}'}{qE\,\lambda(\mathcal{E}')}\right].
\label{eq:PLB}
\end{equation}
Here $\lambda(\mathcal{E})$ is the mean free path for momentum-relaxing collisions.

The LD probability accounts for ballistic motion to an intermediate energy $\mathcal{E}_1$ ($0<\mathcal{E}_1\leq\mathcal{E}$), followed by drift from $\mathcal{E}_1$ to $\mathcal{E}$ with frequent momentum randomization but no energy-relaxing event~\cite{McKenzie1987}:
\begin{equation}
\begin{aligned}
P_{\mathrm{LD}}(\mathcal{E}) &=
\int_{0}^{\mathcal{E}}\frac{d\mathcal{E}_1}{qE\,\lambda(\mathcal{E}_1)}
\exp\!\left[-\int_{0}^{\mathcal{E}_1}\frac{d\mathcal{E}'}{qE\,\lambda(\mathcal{E}')}\right] \\
&\qquad\times
\exp\!\left[-\int_{\mathcal{E}_1}^{\mathcal{E}}\frac{d\mathcal{E}'}{qE\,\lambda_E(\mathcal{E}')}\right].
\end{aligned}
\label{eq:PLD}
\end{equation}
Here $\lambda_E(\mathcal{E})$ is the mean energy relaxation length. In the LD limit, momentum relaxation is rapid compared to energy relaxation, $\lambda(\mathcal{E}) \ll \lambda_E(\mathcal{E})$, allowing carriers to undergo many momentum-relaxing collisions while avoiding an energy-relaxing event long enough to reach high energy. The total probability of a carrier attaining energy $\mathcal{E}$ is given by $P(\mathcal{E}) = P_{\mathrm{LB}}(\mathcal{E}) + P_{\mathrm{LD}}(\mathcal{E})$.

Impact ionization occurs when a carrier attains energy exceeding the ionization threshold $\mathcal{E}_I$. The field-dependent impact ionization coefficient $\alpha(E)$ can then be evaluated from $P(\mathcal{E})$ according to~\cite{Kasap2004}:
\begin{equation}
\alpha(E)=\frac{qE\,P(\mathcal{E}_I)}{\displaystyle\int_{0}^{\mathcal{E}_I} P(\mathcal{E})\,d\mathcal{E}}
\label{eq:LD_alpha}
\end{equation}

If $\lambda(\mathcal{E})$ and $\lambda_E(\mathcal{E})$ are treated as energy-independent, $\lambda(\mathcal{E})=\lambda$ and $\lambda_E(\mathcal{E})=\lambda_E$, and $\lambda \ll \lambda_E$, the LD model admits an analytic low-field approximation~\cite{Kasap2004}. In the regime $\alpha(E)\lambda < 0.1$, or equivalently when $\mathcal{E}_I/(qE\lambda) > 10$, $\alpha(E)$ is given by
\begin{equation}
\alpha(E)=\frac{1}{\lambda_E}\exp\!\left(-\frac{\mathcal{E}_I}{qE\,\lambda_E}\right).
\label{eq:LD}
\end{equation}

Here $1/\lambda_E$ sets the characteristic ionization length scale, while the exponential factor gives the probability of reaching $\mathcal{E}_I$ before an energy-relaxing collision~\cite{Kasap2004,Jandieri2008}.

Eq.~\ref{eq:LD} collapses energy relaxation into a single constant length scale, $\lambda_E$, corresponding to a constant energy-relaxation rate over the interval $0 \le \mathcal{E} \le \mathcal{E}_I$. In a disordered semiconductor, however, the available scattering channels and energy-loss rates generally evolve with carrier energy, so $\lambda_E$ is not expected to remain constant over the full energy range relevant for ionization. An energy-dependent relaxation length, $\lambda_E(\mathcal{E})$, is parameterized in the LD model as~\cite{Kasap2004}:

\begin{equation}
\lambda_E(\mathcal{E})=\lambda_{E0}+ \beta\,\mathcal{E}^{n},
\label{eq:lambdaE_eq10}
\end{equation}
where $\lambda_{E0}$ is the low-energy limit of the energy-relaxation length, $\beta$ sets the magnitude of the energy dependence, and $n$ is a scattering index.

\subsection{Charge transport in a-Se}
\label{sec:transport}
\subsubsection{Transport framework and effective mobility}
\label{sec:transport_framework}

Charge transport in a-Se is governed by its intrinsically disordered atomic network, which gives rise to localized band-tail states and a broad distribution of defect levels within the mobility gap~\cite{mott1979,cohen1969}. Injected or photogenerated carriers move via drift in extended, delocalized states, intermittently interrupted by thermally activated or field-assisted transitions into and out of localized states~\cite{pfister1976,Kasap2015}. As a result, the effective carrier mobility and drift length depend strongly on both temperature and electric field~\cite{Hijazi2014}. These dependences have been documented experimentally from room temperature down to \(166\,\mathrm{K}\) across a wide range of electric fields~\cite{juska,Hijazi2016}.

The effective mobility \(\mu_{\mathrm{eff}}\) is an experimentally inferred transport coefficient, distinct from the intrinsic band mobility. The resulting variation of mobility with temperature and field directly influences the measured current in a-Se devices. The current density is expressed as
\begin{equation}
J(E,T)=q\,p_{\mathrm{free}}(E,T)\,\mu_{\mathrm{eff}}(E,T)\,E,
\label{eq:drift}
\end{equation}
where \(p_{\mathrm{free}}\) is the free-hole density. In a disordered material such as a-Se, both \(p_{\mathrm{free}}\) and \(\mu_{\mathrm{eff}}\) are functions of field and temperature~\cite{Pfister1977,Nenashev2018,tabak1968}. The quantity \(\mu_{\mathrm{eff}}(E,T)\) integrates the effects of trapping, detrapping, scattering, and other field-dependent transport physics. Several candidate descriptions proposed in the literature are used to connect the field and temperature dependences of \(p_{\mathrm{free}}(E,T)\) and \(\mu_{\mathrm{eff}}(E,T)\) in Eq.~\ref{eq:drift} to experimentally measured current, and are grouped into trap-limited drift in extended states, hopping transport between localized states, and field-enhanced emission processes that can involve tunneling.

\subsubsection{Trap-limited drift in extended states}
\label{sec:extended_states}
Charge motion in a-Se is often described using the multiple-trapping model~\cite{Kasap2015}. Carriers spend most of their time immobilized in localized traps and are intermittently released into extended states where they drift until re-captured. Averaging over these trapping cycles yields an effective mobility smaller than the free-carrier mobility, increasing with temperature as thermal energy promotes detrapping. The effective mobility often follows an Arrhenius relation,
\begin{equation}
\mu_{\mathrm{eff}}(T)=\mu_{\infty}\exp\!\left(-\frac{E_a}{k_B T}\right),
\label{eq:Arrhenius}
\end{equation}
where \(E_a\) is the activation energy for release from dominant traps and \(\mu_{\infty}\) is the high-temperature mobility limit. 

Field-assisted lowering of Coulombic trap barriers, known as the Poole--Frenkel (PF) effect, enhances thermal emission from charged traps by reducing the potential barrier that confines carriers, thereby increasing conduction. The reduction in barrier height is expressed as 
\begin{equation}
\Delta \Phi(E) = \beta_{\mathrm{PF}} E^{1/2},
\end{equation}
where \(\beta_{\mathrm{PF}} \!\approx\! \sqrt{q^3/(\pi \varepsilon_0 \varepsilon_r)}\) is the PF coefficient.

The current density under field-assisted emission can then be written as
\begin{equation}
J(E,T) = \sigma_{\mathrm{PF}}\,E \exp\!\left(\frac{\beta_{\mathrm{PF}} E^{1/2}}{k_B T}\right),
\label{eq:PF}
\end{equation}
where \(\sigma_{\mathrm{PF}}\) may be written as \(\sigma_{\mathrm{PF}}=\sigma_{0}\exp(-\Phi_{0}/k_{B}T)\), with \(\Phi_{0}\) the zero-field trap barrier and \(\sigma_{0}\) an effective prefactor that collects the contributions of mobility, carrier concentration, and other factors that determine the low-field conductivity. The PF relation thus predicts an exponential enhancement of conduction with $\sqrt{E}$~\cite{Simmons1971}.

\subsubsection{Hopping transport between localized states}
\label{sec:localized}

When transport proceeds by hopping between localized states, an applied field can assist hops and modify both the temperature and field dependence of conductivity. In the low-field limit, charge transfer between localized states near the Fermi level occurs via phonon-assisted hopping, leading to the three-dimensional Mott variable-range hopping (VRH) relation~\cite{Mott1969}:
\begin{equation}
\sigma(T) = \sigma_{\mathrm{M}} \exp\!\left[-\left(\frac{T_0}{T}\right)^{1/4}\right],
\end{equation}
where \(\sigma(T)\) is the electrical conductivity and \(T_0\) is a characteristic temperature determined by the density of localized states and the localization length. The prefactor \(\sigma_{\mathrm{M}}\) encodes the attempt-to-escape rate, phonon coupling, localization length, the local density of states near the Fermi level, and geometric factors. This relation describes the temperature dependence of conductivity when the applied field is too small to significantly bias hopping, that is when the energy gained over a typical hop, $q E R(T)$, is much less than the thermal energy $k_B T$. A practical criterion for the low-field VRH regime is $q E R(T) \ll k_B T$, where $R(T) \approx a (T_0/T)^{1/4}$ is the typical Mott hop length and $a$ is the localization length~\cite{godet2002,Kuksenkov1998}. Thus a characteristic crossover field is $E_{\mathrm{c}} \!\approx\! (k_B T)/(q a)\,(T/T_0)^{1/4}$. For $E \gtrsim E_{\mathrm{c}}$, field-assisted hopping or other field-dominated mechanisms must be considered.

At higher fields, the electric field tilts the potential landscape, enhancing the probability of hops in the field direction. The field-assisted hopping current can be expressed as~\cite{Hill1971}:  

\begin{equation}
J \propto 
\sinh\!\big[1.03\,qE(\alpha\pi N_i)^{-1/4}(k_B T)^{-5/4}\big]
e^{-C T^{-1/4}}
\label{eq:Hill}
\end{equation}

where \(\alpha = 1/a\)  and \(N_i\) is the localized density of states (DOS) at the Fermi level. The constant \(C\) is related to the Mott characteristic temperature via \(C=T_0^{1/4}\), so \(T_0=C^{4}\) sets the characteristic temperature scale, with an associated energy scale \(k_{B}T_{0}\). The numerical factor 1.03 arises from the combination of geometric terms in the derivation of the hopping expression.

\subsubsection{Field-enhanced emission involving tunneling}
\label{sec:tunneling}
The microscopic mechanism responsible for field-enhanced emission in a-Se remains unresolved, with Poole--Frenkel and thermally assisted tunneling (TAT) models proposed to describe the observed behavior~\cite{Kabir2014}. TAT occurs when an electric field modifies the effective barrier between a localized trap and the transport band, allowing carriers thermally promoted into tunneling-ready states to tunnel the remaining short distance into extended states. A finite thermal population sets the occupation of tunneling-ready states, while the field governs the residual tunneling probability, giving the emission rate an exponential field dependence that strengthens as temperature decreases~\cite{vincent1979}. The resulting current density can be written as
\begin{equation}
J(E,T)=\sigma_0(T)\,E\,\exp\!\left(\frac{q\,a_{T}}{k_B T}\,E\right),
\label{eq:tunnel}
\end{equation}
where \(a_{T}\) is an effective tunneling distance and \(\sigma_0(T)\) is a temperature-dependent prefactor that collects material-specific parameters of the emission and transport process~\cite{vincent1979,Kabir2014}.

Two additional tunneling processes can further contribute to field-enhanced transport in a-Se. Phonon-assisted tunneling (PAT) describes inelastic transitions in which lattice vibrations supply energy to carriers that are otherwise localized below a barrier~\cite{Ganichev2000,Katzenmeyer2010,kleinman1965}. The phonon interaction perturbs the local potential and enables tunneling through a field-narrowed barrier, producing an exponential increase in conductivity with the square of the applied field. The corresponding current density can be written as
\begin{equation}
J = J_{0}\,E\,\exp\!\left(\frac{E^{2}}{E_{c}^{2}}\right),
\label{eq:PAT_revised}
\end{equation}
with the characteristic field
\begin{equation}
E_{c}=\sqrt{\frac{3\,m^{*}\hbar}{q^{2}\tau^{3}}}.
\label{eq:Ec}
\end{equation}
Here \(J_{0}\) is a prefactor that depends on trap density, capture cross section, and attempt-to-escape frequency, \(m^{*}=n\,m_{e}\) is the carrier effective mass expressed as a fraction $n$ of the free-electron mass, \(\hbar\) is the reduced Planck constant, and \(\tau\) is an effective tunneling time that may vary weakly with temperature due to changes in phonon population, trap occupancy, and barrier shape.

Fowler--Nordheim tunneling (FNT) describes elastic quantum tunneling through a field-thinned triangular barrier at an interface or from highly localized states~\cite{Lee2011}. In the usual triangular-barrier approximation, the current density is
\begin{equation}
J = \frac{q^{3}m_{e}}{8\pi h m^{*}}\,\frac{E^{2}}{\varphi}\,
\exp\!\left(-\frac{8\pi\sqrt{2m^{*}}\,\varphi^{3/2}}{3q h E}\right),
\label{eq:FN_revised}
\end{equation}
where \(\varphi\) is the effective barrier height and \(h\) is Planck’s constant. This expression omits image-force and Nordheim function corrections and should be interpreted as the simple triangular-barrier result. Fowler--Nordheim emission is elastic and exhibits negligible explicit temperature dependence.

\section{Results}\label{sec:results}

\subsection{Analysis Definitions}
\label{sec:analysis_defs}

Recorded waveforms from Sec.~\ref{sec:methods} were averaged and fit to extract the peak amplitude response, which was converted to the collected charge per pulse, \(Q_{\mathrm{col}}\), using the charge-sensitive preamplifier gain~\cite{Rooks2026}.

The number of collected holes per pulse, \(n_{\mathrm{h}}\), was calculated as \(n_{\mathrm{h}}=Q_{\mathrm{col}}/q\). The pulse-averaged photocurrent was then calculated as
\begin{equation}
I = q\,n_{\mathrm{h}}\,f,
\end{equation}
where \(f=4\,\mathrm{Hz}\) is the pulse repetition frequency. EQE was calculated as \(\mathrm{EQE}=n_{\mathrm{h}}/N_{\mathrm{ph}}\), where \(N_{\mathrm{ph}}\) is the number of incident photons per pulse~\cite{Rooks2026}.

The measured current is related to drift current density \(J\) by \(I = J A_{\mathrm{eff}}\), where \(A_{\mathrm{eff}}\) is the effective cross-sectional area through which photogenerated charge drifts. A first-order estimate of \(A_{\mathrm{eff}}\) can be obtained from the a-Se/IDE geometry by summing the total gap length within the dot diameter and multiplying by the \(1/e\) absorption depth at \(401\,\mathrm{nm}\), which gives \(A_{\mathrm{eff}}\approx 825\,\mu\mathrm{m}^2\)~\cite{Leiga1968}.

Using Eq.~\ref{eq:drift}, the relation can be written as
\begin{equation}
I = J\,A_{\mathrm{eff}} = q\,p_{\mathrm{free}}(E,T)\,\mu_{\mathrm{eff}}(E,T)\,E\,A_{\mathrm{eff}},
\end{equation}

Dividing both sides by \(E\) yields a useful diagnostic,
\begin{equation}
\frac{I}{E} = q\,p_{\mathrm{free}}(E,T)\,\mu_{\mathrm{eff}}(E,T)\,A_{\mathrm{eff}},
\label{eq:IoverE}
\end{equation}
so that \(I/E\) measures the product of free-hole density and effective mobility scaled by geometric and fundamental constants.

\subsection{Photogeneration Efficiency}
\label{sec:QE}

\begin{figure}
\centering
\includegraphics[width=0.95\columnwidth]{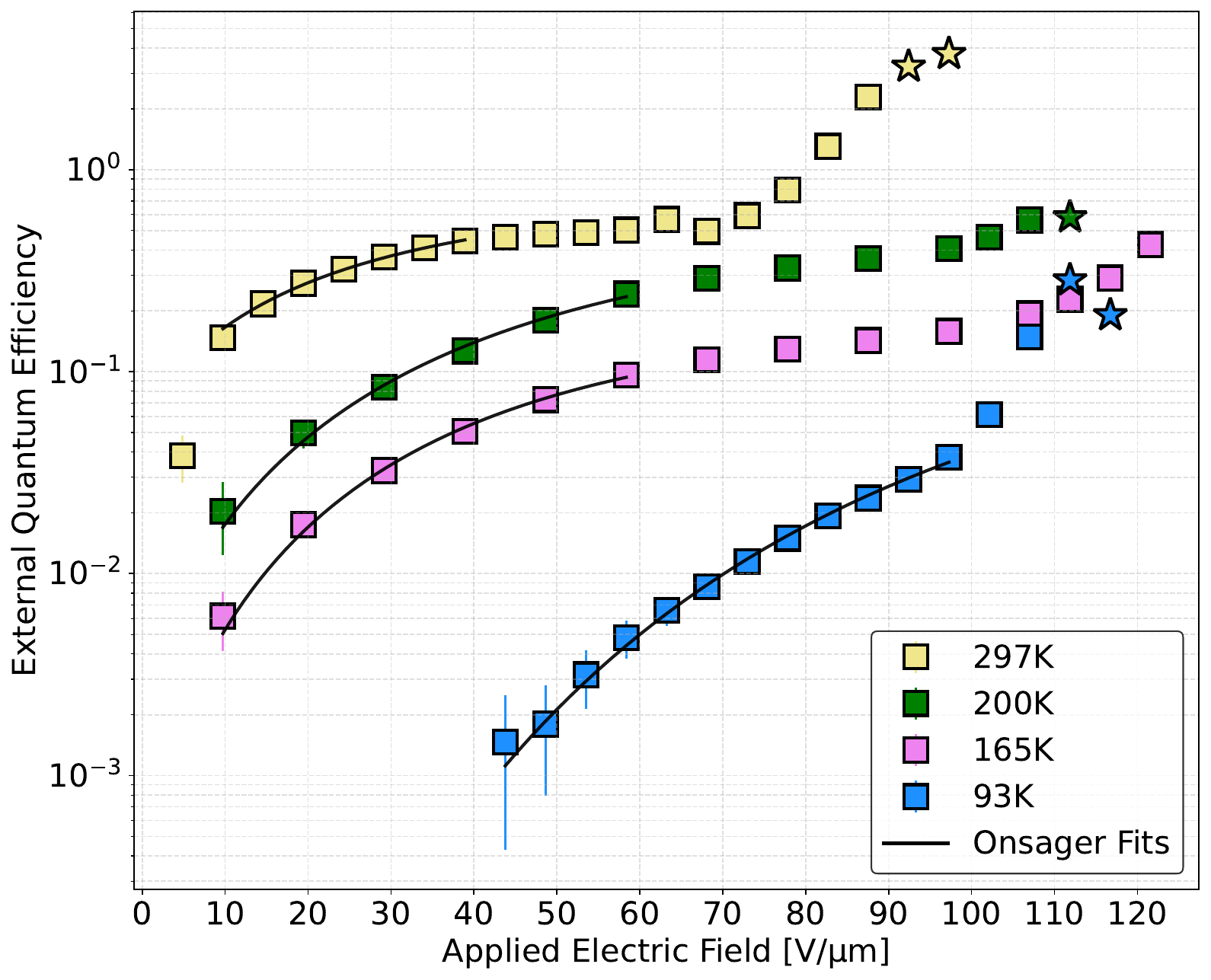}
\caption{Response data expressed as EQE versus applied field for detectors operated at four temperatures with Onsager fits over the field-assisted detrapping region, as discussed in Sec.~\ref{sec:transport_results}. Starred points indicate field screening.}
\label{fig:eqePlot}
\end{figure}

The measured responses are shown in Fig.~\ref{fig:eqePlot} as EQE versus applied field for 93\,K, 165\,K, 200\,K, and 297\,K, each obtained from a separate detector operated at a fixed temperature. At 165\,K, 200\,K, and 297\,K the responses show sublinear growth at low fields, a gradual approach to saturation, and then an abrupt exponential rise marking the onset of avalanche multiplication. At 93\,K the response instead exhibits a slow exponential increase across the low-field region without clear saturation before avalanche onset. At the highest fields the curves exhibit a decrease in \(\mathrm{d}EQE/\mathrm{d}E\) and a negative curvature \(\mathrm{d}^2EQE/\mathrm{d}E^2<0\), identified in Fig.~\ref{fig:eqePlot} by starred markers and consistent with field screening in which injected carriers accumulate in the transport path and reduce the effective lateral field.

To assess whether the field dependence of the response followed the expected behavior of geminate pair dissociation, the EQE data were fit using the Onsager model from Sec.~\ref{sec:Theory}, treating the overall scale at each temperature as a single fitted prefactor:
\begin{equation}
\mathrm{EQE}(E,T)=\xi_{\mathrm{eff}}(T)\,\eta_{\mathrm{Ons}}(E,T;r_{0}),
\label{eq:EQE_ons_fit}
\end{equation}
where \(\eta_{\mathrm{Ons}}(E,T;r_{0})\) is given by Eqs.~\ref{eq:yip1}--\ref{eq:yip2}. Because \(\mathrm{EQE}=\eta\,\xi\) and the Onsager expression specifies the field and temperature dependence of \(\eta\) up to the field-independent factor \(\eta_{0}\), the fitted \(\xi_{\mathrm{eff}}(T)\) should be interpreted as an effective scale factor that absorbs \(\eta_{0}\) together with collection losses over the fitted interval, rather than as a determination of the intrinsic material parameter \(\eta_{0}\), which is commonly taken as unity for a-Se~\cite{Pai1975}.

The Onsager model reproduced the overall field dependence but yielded \(r_{0}\) values smaller than the 7.0\,nm thermalization length reported for 400\,nm excitation between 223\,K and 294\,K~\cite{Hijazi2016,Pai1975}. 
Fits were restricted to the field-assisted detrapping region, which is defined later in Sec.~\ref{sec:transport_results}. Figure~\ref{fig:eqePlot} shows the Onsager fits, and the corresponding parameters are listed in Table~\ref{tab:Onsager_results}.

\begin{table}
\setlength{\tabcolsep}{10pt}
\centering
\begin{tabular}{cccc}
\hline
\multicolumn{4}{c}{\textbf{Onsager Fit Parameters}} \\
\hline
\makecell{Temp \\ (K)} &
\makecell{$\xi_{\mathrm{eff}}$} &
\makecell{$r_{0}$ \\ (nm)} &
$\chi^{2}/\mathrm{dof}$ \\
\hline
297 & $0.91 \pm 0.04$ & $3.07 \pm 0.14$ & 0.697 \\
200 & $0.75 \pm 0.03$ & $2.15 \pm 0.06$ & 0.627 \\
165 & $0.26 \pm 0.01$ & $2.31 \pm 0.04$ & 0.965 \\
 93 & $0.37 \pm 0.01$ & $1.42 \pm 0.01$ & 0.362 \\
\hline
\end{tabular}

\caption{Fits used a fixed relative permittivity \(\varepsilon_{r}=6.7\) for a-Se. For comparison, an optical thermalization length of \(7.0\,\mathrm{nm}\) was reported for 400\,nm excitation~\cite{Pai1975}.}

\label{tab:Onsager_results}
\end{table}

\subsection{Avalanche Multiplication}
\label{sec:avalanche}

\begin{figure}
\centering
\includegraphics[width=\columnwidth]{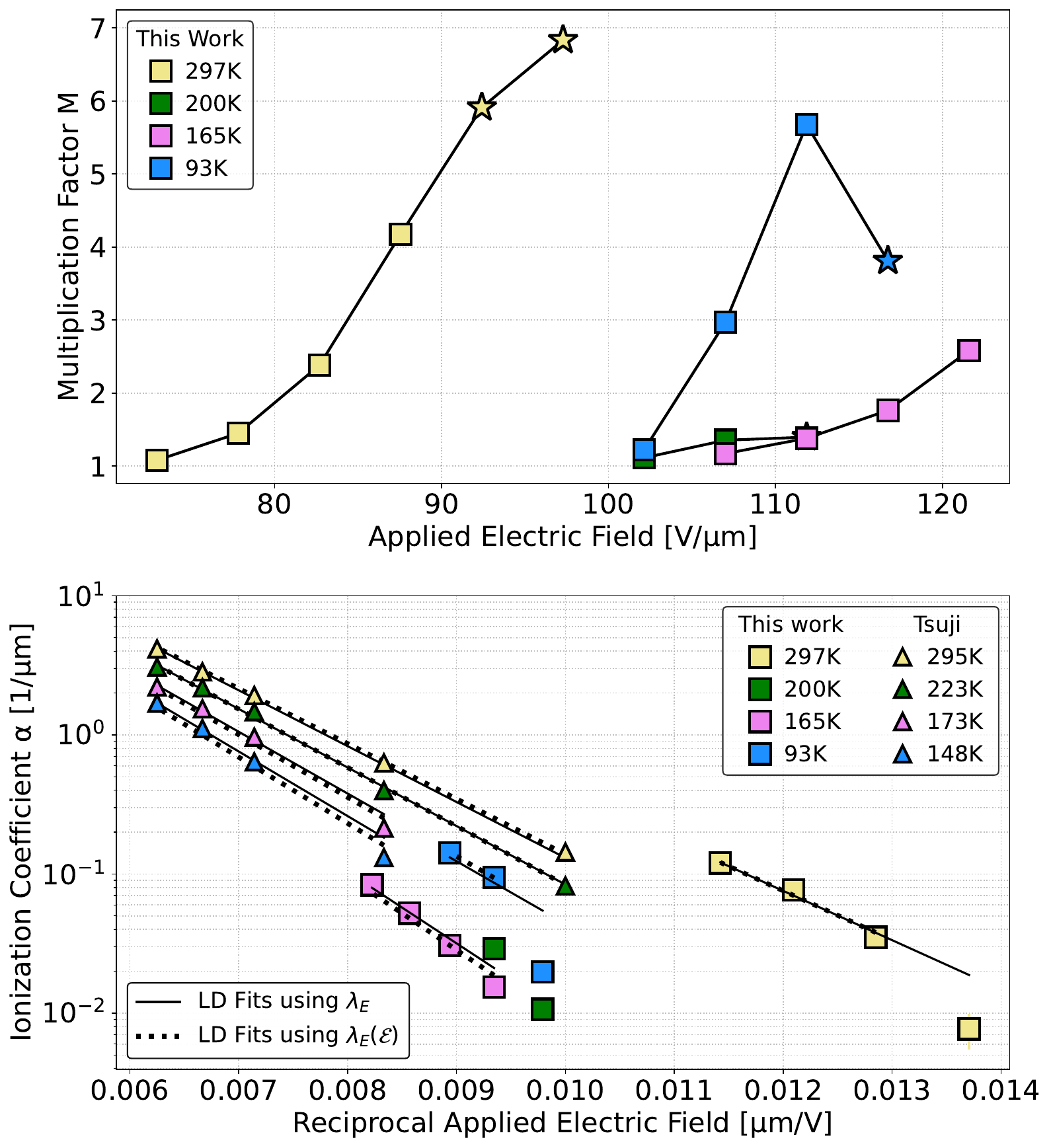}
\caption{Top: Multiplication factor calculated from EQE response data in the avalanche regime. Bottom: Ionization coefficients from this work together with data extracted from Tsuji~\cite{Tsuji1989}, both fit with the LD model for comparison.}
\label{fig:LDPlot}
\end{figure}


\setlength{\tabcolsep}{6pt}
\begin{table}
\centering
\begin{tabular}{@{\hspace{0.5cm}}c@{\hspace{1cm}}c@{\hspace{1cm}}c@{\hspace{1cm}}c@{}}

\hline
\makecell{Temp \\ (K)} & \makecell{$\lambda_E$ \\ (nm)} & \makecell{$E_I$ \\ (eV)} & $R^{2}$ \\
\hline
\multicolumn{4}{c}{\textbf{Energy-independent Lucky-drift Fits to this work}} \\
\hline
297 & 0.74 & $0.604 \pm 0.003$   & 0.975 \\
200 & \textemdash & \textemdash & \textemdash \\
165 & 0.74 & $0.876 \pm 0.004$ & 0.977 \\
 93 & 0.74 & $0.765 \pm 0.013$ & 0.818 \\
\hline
\multicolumn{4}{c}{\textbf{Energy-independent Lucky-drift Fits to Tsuji}} \\
\hline
295 & 0.74 & $0.684 \pm 0.002$ & 0.999 \\
223 & 0.74 & $0.717 \pm 0.002$ & 0.995 \\
173 & 0.74 & $0.757 \pm 0.002$ & 0.994 \\
148 & 0.74 & $0.791 \pm 0.001$ & 0.997 \\
\hline
\end{tabular}
\caption{Fit parameters for the energy-independent relaxation-length model, Eq.~\ref{eq:LD}, from this work and from fits to the data of Tsuji~\cite{Tsuji1989}. For all fits in this table, $\lambda_E$ was fixed at $0.74\,\mathrm{nm}$. The $200\,\mathrm{K}$ parameters are not reported due to limited avalanche-region data.}

\label{tab:LD_results}
\end{table}


\setlength{\tabcolsep}{8pt}
\begin{table}
\centering
\begin{tabular}{cccc}
\hline
\makecell{Temp \\ (K)} &
\makecell{$\beta$ \\ (nm\,eV$^{-1}$)} &
\makecell{$\lambda_E(\mathcal{E}_I)$ \\ (nm)} &
$R^{2}$ \\
\hline
\multicolumn{4}{c}{\textbf{Energy-dependent Lucky-drift Fits to this work}} \\
\hline
297 & $1.231 \pm 0.017$ & $1.931 \pm 0.017$ & 0.984 \\
200 & \textemdash & \textemdash & \textemdash \\
165 & $0.252 \pm 0.012$ & $0.952 \pm 0.012$ & 0.948 \\
 93 & $0.5765 \pm 0.0014$ & $1.2765 \pm 0.0014$ & 0.999 \\
\hline
\multicolumn{4}{c}{\textbf{Energy-dependent Lucky-drift Fits to Tsuji}} \\
\hline
295 & $0.862 \pm 0.005$ & $1.562 \pm 0.005$ & 0.998 \\
223 & $0.716 \pm 0.009$ & $1.416 \pm 0.009$ & 0.995 \\
173 & $0.561 \pm 0.020$ & $1.261 \pm 0.020$ & 0.987 \\
148 & $0.445 \pm 0.021$ & $1.145 \pm 0.021$ & 0.977 \\
\hline
\end{tabular}
\caption{Fit parameters for the energy-dependent relaxation-length model, Eq.~\ref{eq:lambdaE_eq10}, with $n=1$. The values of $\lambda_E(\mathcal{E}_I)$ are reported at fixed $\mathcal{E}_I=1.0\,\mathrm{eV}$. In these fits, $\lambda=0.30\,\mathrm{nm}$ and $\lambda_{E0}=0.70\,\mathrm{nm}$ were fixed, and $\mathcal{E}_I$ was fixed at $1.0\,\mathrm{eV}$.}

\label{tab:LD_energydep_results}
\end{table}

At higher fields, the EQE response exhibits an abrupt exponential increase that marks the onset of avalanche multiplication. To quantify this regime, the data at each temperature were normalized at the avalanche point identified from the transport analysis later in Sec.~\ref{sec:transport_results}, such that $M = 1$ at the onset point and all higher field values represent charge multiplication.

The resulting gain curves are shown in Fig.~\ref{fig:LDPlot} (top) and were transformed using Eq.~\ref{eq:Mavg} to obtain the impact ionization coefficients (IIC), presented in Fig.~\ref{fig:LDPlot} (bottom) together with literature values from Tsuji~\cite{Tsuji1989}. These data are included because they represent the only published measurements of avalanche in a-Se at cryogenic temperatures, obtained over a similar temperature range and using 400\,nm optical excitation, making them a natural benchmark for comparison, with device differences discussed in Sec.~\ref{sec:discussion}.

Both data sets were fit with the energy-independent low-field LD form in Eq.~\ref{eq:LD} after verifying that $\alpha(E)\lambda<0.1$ over the fit range. The mean free path for momentum relaxation was fixed at $\lambda=0.30\,\mathrm{nm}$, on the order of one atomic spacing in a-Se~\cite{Kasap2004}. Because Eq.~\ref{eq:LD} yields strongly correlated estimates of $(\mathcal{E}_I,\lambda_E)$, $\lambda_E$ was fixed to regularize the fit and enable direct comparisons across temperatures, while $\mathcal{E}_I$ was allowed to vary. 

As a consistency check, the 295\,K subset of the Tsuji IIC data used here was refit with the same LD expression used by Kasap, fixing $\lambda_E=0.74\,\mathrm{nm}$ and allowing $\mathcal{E}_I$ to vary~\cite{Kasap2004,Tsuji1989}. This procedure yields $\mathcal{E}_I = 0.684 \pm 0.002$\,eV, within 3.6\% of the $\mathcal{E}_I \approx 0.66$\,eV reported by Kasap from a larger room-temperature data set~\cite{Kasap2004}. With $\lambda_E$ held fixed, the remaining Tsuji IIC data show a systematic increase in $\mathcal{E}_I$ with decreasing temperature, as summarized in Table~\ref{tab:LD_results}, and the same fitting constraints were then applied to the IIC data from this work. The fitted $\mathcal{E}_I$ values from this work were non-monotonic with temperature, in contrast to the Tsuji trend. This energy-independent LD form describes the Tsuji IIC data well across the full temperature range, as reflected by the $R^{2}$ values in Table~\ref{tab:LD_results}. The 93\,K IIC data from this work are less well described.

To allow for non-constant energy relaxation, Eqs.~\ref{eq:PLB}--\ref{eq:LD_alpha} were evaluated numerically using the energy-dependent relaxation-length model of Eq.~\ref{eq:lambdaE_eq10}. The momentum-relaxation length was fixed at $\lambda=0.30\,\mathrm{nm}$. The threshold energy was fixed at $\mathcal{E}_I=1.0\,\mathrm{eV}$, assuming that ionization can proceed from midgap localized states in a-Se. The scattering index was set to $n=1$, which provides the simplest energy dependence, and the low-energy limit was held fixed at $\lambda_{E0}=0.70\,\mathrm{nm}$ to reduce correlations with $\beta$ and enable direct comparisons across temperatures and data sets. This $\lambda_{E0}$ value was taken from the fitted results reported in~\cite{Kasap2004}. The corresponding $\lambda_E(\mathcal{E}_I)$ values were computed from the fitted $\beta$ and are reported in Table~\ref{tab:LD_energydep_results}.

Using the same fixed parameters as Kasap, the 295\,K subset of the Tsuji IIC data yields $\lambda_E(\mathcal{E}_I)=1.562 \pm 0.005$\,nm, within 5.3\% of the reported $\lambda_E(\mathcal{E}_I)=1.65$\,nm~\cite{Kasap2004}. Across the remaining Tsuji temperatures, $\beta$ and $\lambda_E(\mathcal{E}_I)$ decrease systematically with decreasing temperature, as summarized in Table~\ref{tab:LD_energydep_results}, and the same fitting constraints were applied to the IIC data from this work. For the 297\,K and 93\,K data from this work, the lowest-field IIC points were excluded because they yielded $\beta$ values that implied non-physical $\lambda_E(\mathcal{E}_I)$ and degraded the fit quality. The IIC data from this work yield larger values at 297\,K and a non-monotonic temperature dependence at lower temperatures, with the smallest $\beta$ and $\lambda_E(\mathcal{E}_I)$ at 165\,K and intermediate values at 93\,K. Differences between the temperature dependences inferred from the Tsuji IIC data and the IIC data from this work are addressed in Sec.~\ref{sec:discussion}.

\subsection{Charge transport regimes and model fits}
\label{sec:transport_results}

\begin{figure}
\centering
\includegraphics[width=\columnwidth]{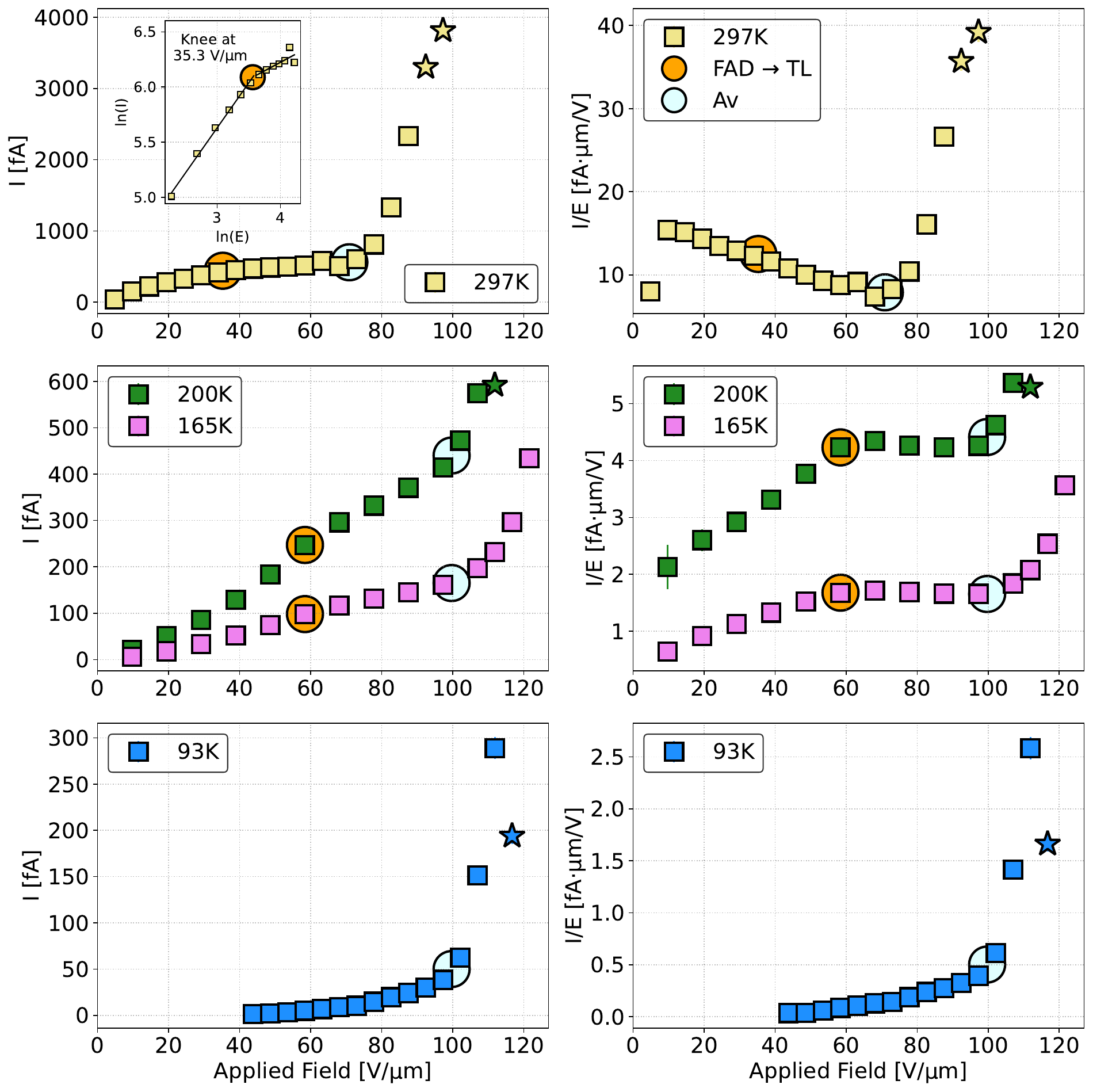}
\caption{Measured photocurrent \(I\) (left) and diagnostic \(I/E\) (right) as functions of applied field for detectors operated at 297\,K, 200\,K, 165\,K, and 93\,K. The plots illustrate transitions between FAD, TL and AV regimes. The inset highlights the knee at 35.3~V/\textmu m in the 297\,K data.}
\label{fig:mu}
\end{figure}

Measured photocurrent \(I\) and the diagnostic \(I/E\) are shown in Fig.~\ref{fig:mu}. Charge transport refers to the portion of the response preceding impact ionization, while a sharp exponential increase marks the transition to avalanche (AV). Data from the AV region are also shown to illustrate the continuation of the response and identify the transition to carrier multiplication.

A rising $I/E$ with field indicates an increasing transport product \(p_{\mathrm{free}}\mu_{\mathrm{eff}}\), reflecting either enhanced free-carrier density through field-assisted detrapping (FAD) or an increase in effective mobility. This trend appears at 200\,K and 165\,K in Fig.~\ref{fig:mu}, from the lowest applied field of 9.7\,V/\textmu m up to the knee at 58.4\,V/\textmu m. When $I/E$ becomes weakly field-dependent, the transport product no longer increases appreciably with field, consistent with transport-limited (TL) behavior. This region is clearly observed at 200\,K and 165\,K, starting at the knee near 58.4\,V/\textmu m and continuing up to just before AV onset at 99.7\,V/\textmu m.

A decreasing $I/E$ with field signifies a reduction in the transport product, consistent with additional transport suppression or a field-dependent decrease in the available mobile population. This behavior is evident in the 297\,K data, beginning near 9.7\,V/\textmu m and extending continuously to the onset of AV.

At 93\,K no TL plateau is observed in the pre-avalanche window. Instead \(I/E\) grows rapidly with field, consistent with strongly suppressed thermal detrapping so that the mobile carrier population remains small until field-assisted mechanisms such as hopping or phonon-assisted tunneling enhance carrier release. Strong field sensitivity in \(p_{\mathrm{free}}(E)\) at low temperature therefore prevents the emergence of a constant \(I/E\) region prior to AV.

\begin{figure}
\centering
\includegraphics[width=0.95\columnwidth]{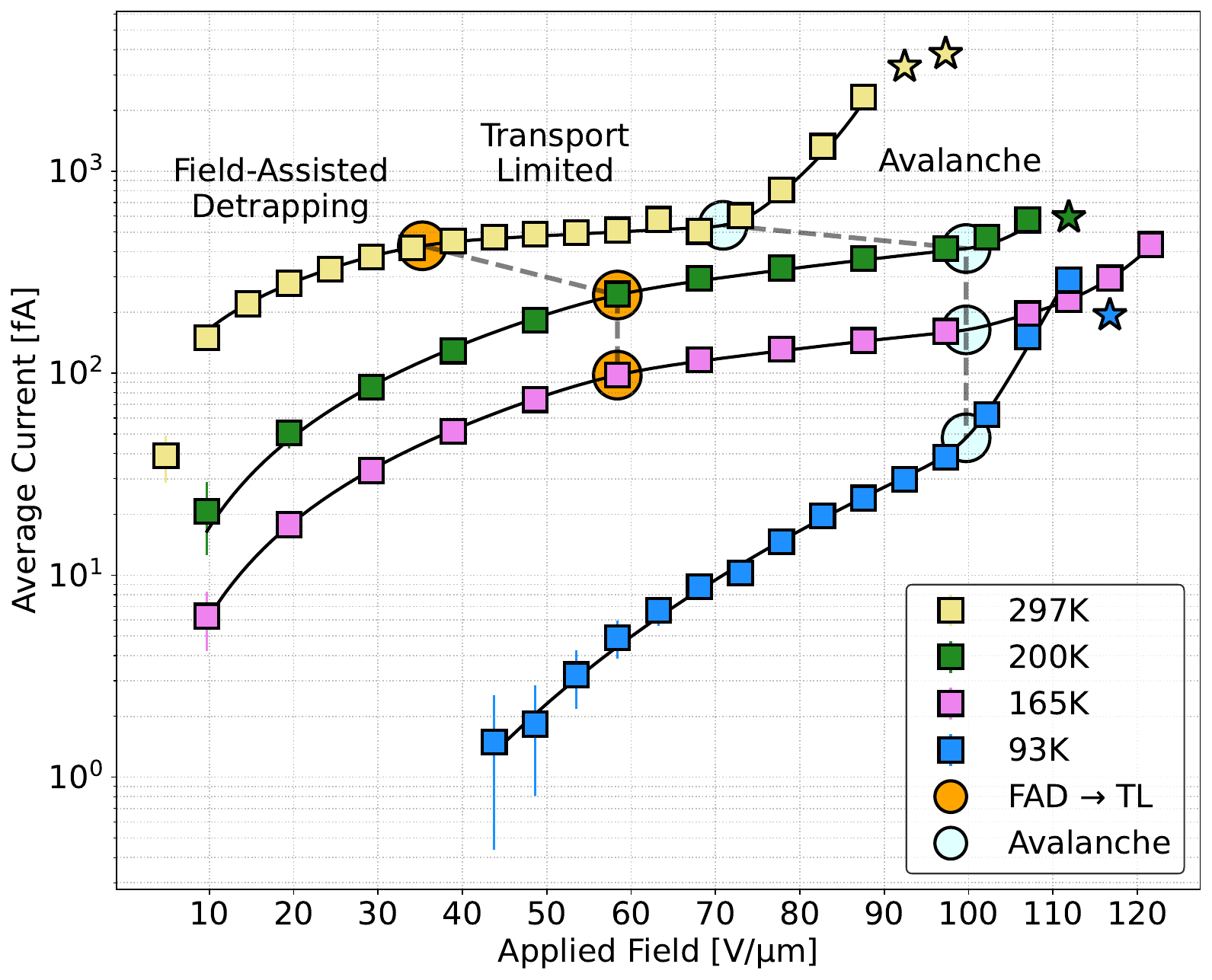}

\caption{Measured photocurrent as a function of applied field for detectors operated at 297\,K, 200\,K, 165\,K, and 93\,K. Data are shown with a hybrid empirical description linking FAD, TL, and AV regimes. Markers indicate the transition fields. The 200\,K AV curve is shown as a guide to the eye.}

\label{fig:QPlot}
\end{figure}

\subsubsection{Phenomenological segmentation}
\label{sec:phen_seg}

The four response curves are shown together in Fig.~\ref{fig:QPlot}, plotted as $I$ vs.\ $E$. Each response was modeled using a hybrid description combining separate forms for the FAD, TL, and AV regions. The FAD and TL segments were fit with power laws, \(I(E)=A\,E^{p}\). The AV region was described using \(I(E)=A\,\exp\!\big[(E/E_{0})^{b}\big]\). At 200\,K, the number of points above AV onset is insufficient to independently constrain \(E_{0}\) and \(b\). Therefore, the 200\,K AV curve shown in Fig.~\ref{fig:QPlot} is shown only as a guide to the eye, and no AV fit parameters are reported for that temperature. The corresponding fitted parameters are summarized in Table~\ref{tab:fitparams_all}, except where noted. Parameters \(A\), \(p\), \(E_0\), and \(b\) represent the scaling constant, power-law exponent, characteristic field, and field exponent, respectively.

The empirical fits in Fig.~\ref{fig:QPlot} display a consistent phenomenology across temperature that maps directly onto the \(I/E\) diagnostic. At 200\,K and 165\,K the responses exhibit three operative intervals. At low applied fields \(\leq 58.4\)\,V/\textmu m the fitted exponent \(p > 1\) implies \(I/E\) increases with field, indicating strong field enhancement of the transport product. At intermediate pre-avalanche fields \(\geq 58.4\)\,V/\textmu m the exponent approaches unity, so that \(I/E\) is approximately constant. At 200\,K and 165\,K, the TL interval manifests as this \(I/E\) plateau, indicating that increasing field no longer produces substantial growth in \(p_{\mathrm{free}}\mu_{\mathrm{eff}}\), consistent with saturation of the transport product over that interval. At 99.7\,V/\textmu m the response enters the AV interval, marked by an abrupt super-exponential rise in current.

At 297\,K the behavior differs. The first measured point lies below its neighbors, as seen in the \(I/E\) plot in Fig.~\ref{fig:mu}, and is treated as incomplete collection. Excluding that point, the low-field region \(\leq 35.3\)\,V/\textmu m is sublinear with \(p<1\), implying that \(I/E\) decreases with field. The observed decrease indicates that as field increases the transport product \(p_{\mathrm{free}}\mu_{\mathrm{eff}}\) does not grow and may even decline. No clear \(I/E\) plateau is observed at 297\,K. Instead, the boundary between the two pre-avalanche intervals (FAD and TL) is identified empirically using the visually apparent knee highlighted in the inset of Fig.~\ref{fig:mu} and the corresponding change in fitted exponent from \(p=0.793\pm0.026\) below the knee to \(p=0.308\pm0.105\) above the knee in Fig.~\ref{fig:QPlot}. Above this knee, the current becomes more sublinear with field, followed by the transition into the AV region at 70.9\,V/\textmu m.

With no TL region visible in the 93\,K response, the large fitted exponent in the pre-avalanche region \(\leq 99.7\)\,V/\textmu m indicates strong field sensitivity dominates transport and the response does not pass through a region of constant \(I/E\) before the onset of multiplication at 99.7\,V/\textmu m.

\setlength{\tabcolsep}{2.4pt}
\begin{table*}
\centering
\begin{tabular}{|c!{\vrule width 0.8pt}ccc!{\vrule width 0.8pt}ccc!{\vrule width 0.8pt}cccc|}
\hline
\multicolumn{1}{|c!{\vrule width 0.8pt}}{\textbf{Temp}} &
\multicolumn{3}{c!{\vrule width 0.8pt}}{\textbf{Field-Assisted Detrapping}} &
\multicolumn{3}{c!{\vrule width 0.8pt}}{\textbf{Transport-Limited}} &
\multicolumn{4}{c|}{\textbf{Avalanche}} \\
\hline
\rule{0pt}{2.5ex}
[K] & A [fA] & p & R$^2$ & A [fA] & p & R$^2$ & A [fA] & E$_0$ [V\!\textmu m$^{-1}$] & b & R$^2$ \\
\hline
297 & $25.7 \pm 2.1$ & $0.793 \pm 0.026$ & 0.995 & $147 \pm 61$ & $0.308 \pm 0.105$ & 0.635 & $138 \pm 44$ & $76.9 \pm 4.3$ & $5.8 \pm 1.4$ & 0.999 \\
200 & $0.876 \pm 0.098$ & $1.373 \pm 0.033$ & 0.998 & $4.42 \pm 0.59$ & $0.992 \pm 0.031$ & 0.997 & --- & --- & --- & --- \\
165 & $0.189 \pm 0.006$ & $1.535 \pm 0.009$ & 0.999 & $1.99 \pm 0.28$ & $0.997 \pm 0.030$ & 0.997 & $139 \pm 22$ & $120.4 \pm 0.2$ & $10.0 \pm 0.4$ & 1.000 \\
93 & $(2.34 \pm 0.83) \times10^{-7}$  & $4.13 \pm 0.08$ & 0.996 & --- & --- & --- & $1.01 \pm 0.62$ & $71 \pm 36$ & $3.7 \pm 2.8$ & 0.991 \\

\hline
\end{tabular}

\caption{Fitted parameters for empirical models corresponding to the field-assisted detrapping, transport-limited, and avalanche regions shown in Fig.~\ref{fig:QPlot}. Avalanche parameters are not reported for 200\,K.}

\label{tab:fitparams_all}
\end{table*}

\begin{table}
\centering
\begin{tabular}{c c c}
\hline
Temp & FAD$\rightarrow$TL & AV \\
{[K]} & {[V/\textmu m]} & {[V/\textmu m]} \\
\hline
297 & $35.3 \pm 2.5$ & $70.9 \pm 2.5$ \\
200 & $58.4 \pm 5$ & $99.7 \pm 2.5$ \\
165 & $58.4 \pm 5$ & $99.7 \pm 2.5$ \\
93  & ---             & $99.7 \pm 2.5$ \\
\hline
\end{tabular}
\caption{Field boundaries for the transition from FAD to TL and AV regions at each temperature.}
\label{tab:field_bounds}
\end{table}

The field boundaries highlighted in Fig.~\ref{fig:QPlot} and listed in Table~\ref{tab:field_bounds} shifted to higher fields upon cooling. The FAD$\rightarrow$TL knee shifted from $35.3 \pm 2.5$\,V/\textmu m at 297\,K to $58.4 \pm 5$\,V/\textmu m at 200\,K and 165\,K, while AV onset rose from $70.9 \pm 2.5$\,V/\textmu m at 297\,K to $99.7 \pm 2.5$\,V/\textmu m at 200\,K, 165\,K, and 93\,K.

\subsubsection{Candidate transport mechanisms}
\label{sec:cand_mech}
\begin{figure}
\centering
\includegraphics[width=\columnwidth]{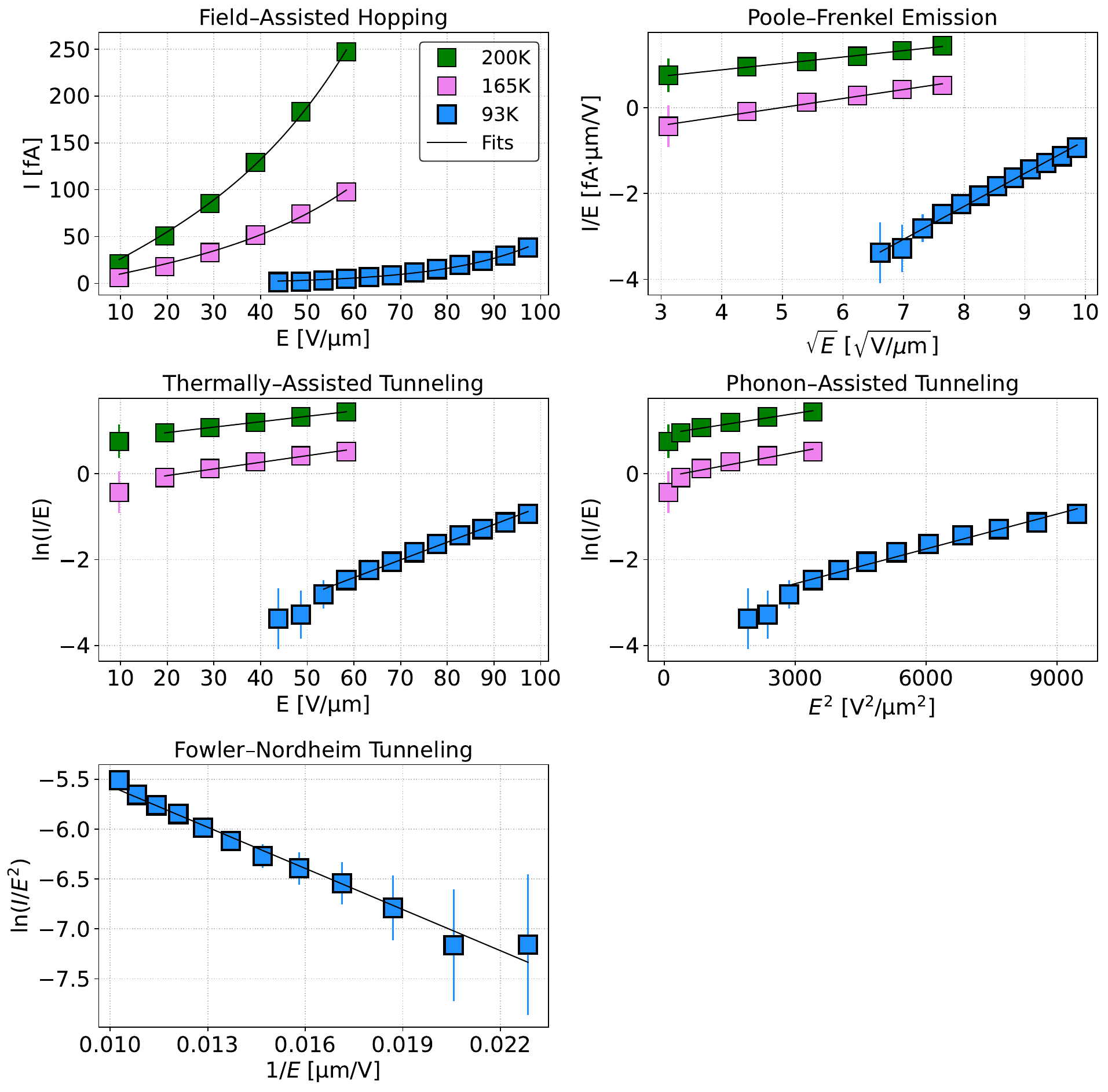}
\caption{Summary of charge transport fits applied to the 200\,K, 165\,K, and 93\,K data. Each model describes a distinct conduction mechanism: field-assisted hopping, Poole--Frenkel emission, thermally assisted tunneling, phonon-assisted tunneling and Fowler--Nordheim tunneling. Solid lines show fitted models.}

\label{fig:summary}
\end{figure}

Several charge transport models were tested against the FAD regions of the 200\,K, 165\,K, and 93\,K responses. The corresponding fits and extracted parameters are summarized in Fig.~\ref{fig:summary} and Table~\ref{tab:summary_params}. While the 297\,K response was analyzed for completeness, the resulting fits were inconsistent with the expected field dependence and were therefore excluded. 

\paragraph*{Field-assisted hopping.}
Field-assisted hopping describes hopping conduction in disordered materials, where carriers move by phonon-assisted transitions between localized states and the applied field favors hops in the drift direction~\cite{Hill1971}. The FAD-region current was fit using Eq.~\ref{eq:Hill} in the form
\vspace{-0.3em}
\begin{equation}
I = A\,\sinh\!\big[1.03\,qE\,B\,(k_B T)^{-5/4}\big]\,e^{-C\,T^{-1/4}}.
\label{eq:Hill2}
\end{equation}

Here \(A\) is a scaling prefactor that absorbs geometric and proportionality constants. The parameter \(B=(\alpha\pi N_i)^{-1/4}\) combines the localization parameter \(\alpha=1/a\), with \(a\) the localization length, and the localized DOS \(N_i\). The hopping coefficient \(C\), introduced in Sec.~\ref{sec:transport} and treated as a fit value here, sets the strength of the Mott-like \(T^{-1/4}\) dependence.

Fits showed consistency across temperature. The localization length is expected to be of order 1~nm, with 0.28~nm reported by~\cite{Kasap2004}. The DOS for a-Se doped with 0.2\% arsenic was reported as \(N_i = 4.14\times10^{46}\ \mathrm{J^{-1}\,m^{-3}}\)~\cite{Serdouk2015}. Using \(a = 0.28\)~nm and the \(N_i\) above yields \(B_{\mathrm{theory}} \approx 6.82\times10^{-15}\ \mathrm{J^{1/4}\!\cdot m}\). Comparing this value to the fitted \(B\) gives relative differences of approximately 44.4\% at 200\,K, 54.6\% at 165\,K and 65.1\% at 93\,K.

The fitted \(C\) values were converted to the Mott characteristic temperature using \(T_0=C^4\). Based on representative values of \(N_i\) and \(a\) reported in the literature, \(T_0\) is expected to fall in the range \(10^{6}\)--\(10^{8}\)~K for amorphous semiconductors~\cite{Kuksenkov1998,Hill1971,godet2002,mott1979}. For the three measured temperatures, \(T_0 \approx 3\times10^{7}\)~K is obtained.

\paragraph*{Poole--Frenkel emission.}
This model treats the FAD-region field dependence as field-assisted thermal emission from charged traps~\cite{Simmons1971}. The PF expression in Eq.~\ref{eq:PF} was fit using the linearized form
\begin{equation}
\ln\!\left(\frac{I}{E}\right)=\ln(A\,\sigma_{\mathrm{PF}})+\frac{\beta_{\mathrm{PF}}}{k_B T}\,E^{1/2},
\end{equation}
where the slope equals \(\beta_{\mathrm{PF}}/(k_B T)\) and the intercept corresponds to \(\ln(A\,\sigma_{\mathrm{PF}})\). All fits showed excellent agreement with the data. However, the extracted \(\beta_{\mathrm{PF}}\) values are consistently an order of magnitude smaller than the theoretical value calculated from \(\beta_{\mathrm{PF}}=\sqrt{q^{3}/(\pi\varepsilon_{0}\varepsilon_{r})}\). Using \(\varepsilon_{r}=6.7\) for a-Se gives \(\beta_{\mathrm{PF,th}}=4.69\times10^{-24}\,\mathrm{J\,m^{1/2}\,V^{-1/2}}\). Solving for \(\varepsilon_{r}\) using the measured \(\beta_{\mathrm{PF}}\) instead yields unrealistically large values that are physically implausible. The Hartke three-dimensional PF formulation was also tested, but yielded no improvement and produced \(\beta_{\mathrm{PF}}\) values comparable to those obtained from the standard PF analysis~\cite{Hartke}.

\paragraph*{Thermally--assisted tunneling.}
TAT treats the FAD-region field dependence as thermally promoted escape from localized states followed by field-driven tunneling into transport states, giving an exponential increase of the current with field~\cite{Kabir2014,vincent1979}. The data were fit using Eq.~\ref{eq:tunnel} in the linearized form
\vspace{-0.3em}
\begin{equation}
\ln\!\left(\frac{I}{E}\right)=\ln\!\big(A\,\sigma_0(T)\big)+\frac{q\,a_{T}}{k_B T}\,E,
\label{eq:tunnel_fit}
\end{equation}
\noindent where the slope corresponds to \(q\,a_{T}/(k_B T)\) and the intercept equals \(\ln\!\big(A\,\sigma_0(T)\big)\).

The lowest-field points, which deviated from linear behavior, were excluded. All fits showed strong linearity with \(R^2>0.98\). The tunneling distance was obtained from the slope through \(a_{T}=\mathrm{slope}\,k_B T/q\). The resulting values are 0.22\,nm at 200\,K, 0.22\,nm at 165\,K, and 0.33\,nm at 93\,K, smaller than the 1.2\,nm effective tunneling length reported for vertically structured a-Se devices~\cite{Hijazi2015_2}.

\paragraph*{Phonon--assisted tunneling.}
PAT treats the FAD-region field dependence as field-enhanced emission from localized states, and inelastic phonon interactions provide the energy exchange that enables tunneling~\cite{Ganichev2000}. The PAT expression in Eq.~\ref{eq:PAT_revised} was fit using the linearized form
\begin{equation}
\ln\!\left(\frac{I}{E}\right)=\ln(AJ_{0})+s\,E^{2},
\end{equation}
where
\begin{equation}
s=\frac{1}{E_{c}^{2}} \quad\text{and}\quad
E_{c}=\sqrt{\frac{3\,m^{*}\hbar}{q^{2}\tau^{3}}},
\end{equation}
and the intercept equals \(\ln(AJ_0)\). The lowest-field points, which deviated from linear behavior, were excluded. All fits showed reasonable linearity with \(R^{2}\) values of 0.978, 0.919 and 0.970 for 200~K, 165~K and 93~K respectively, though the agreement is somewhat poorer than for the other models. If the effective mass \(m^{*}\) is known the tunneling time \(\tau\) can be obtained from the fitted \(E_c\) via
\begin{equation}
\tau=\left(\frac{3\,m^{*}\hbar}{q^{2}E_{c}^{2}}\right)^{1/3}.
\end{equation}
Assuming \(m^*=m_e\), the characteristic field obtained from the slope \(E_c\) = 79.4 V/\textmu m, 72.4 V/\textmu m, and 60.8 V/\textmu m, which correspond to effective tunneling times of \(12.1\ \text{fs},\ 12.9\ \text{fs},\) and \(14.5\ \text{fs}\) at \(200\ \text{K},\ 165\ \text{K},\) and \(93\ \text{K}\), respectively. Varying \(m^*\) between \(0.1\,m_e\) and \(2.0\,m_e\) changes the inferred tunneling times modestly, but \(\tau\) remains of order 10 fs, consistent with PAT observed in other semiconductor materials~\cite{Ganichev2000}.

\paragraph*{Fowler--Nordheim tunneling.}
FNT describes elastic, field-driven tunneling through a field-thinned triangular barrier~\cite{Lee2011}. The data were analyzed using the linearized form of Eq.~\ref{eq:FN_revised},
\begin{equation}
\ln\!\left(\frac{I}{E^{2}}\right)
    = \ln(AC) + \Lambda\,\frac{1}{E},
\end{equation}
where
\begin{equation}
C = \frac{q^{3}m_{e}}{8\pi h m^{*}\varphi},
\qquad
\Lambda = -\frac{8\pi\sqrt{2m^{*}}\,\varphi^{3/2}}{3 q h}.
\end{equation}
Only the 93\,K data exhibited sufficient linearity for fitting, with \(R^{2}=0.98\). Interpreting the fitted slope to extract an effective barrier height and assuming \(m^{*}=m_{e}\) gives \(\varphi \approx 0.074\,\mathrm{eV}\), which is too low to be physically realistic.

\begin{table*}
\centering
\begin{tabular}{lccc}
\hline
\hline
\multicolumn{4}{c}{\textbf{Field-Assisted Hopping}} \\
\hline
Parameter & 200~K & 165~K & 93~K \\
\hline
\noalign{\vskip 1pt}
A [Amps] & $(5.47 \pm 1.00)\times10^{-5}$ & $(7.26 \pm 0.79)\times10^{-5}$ & $(4.59 \pm 0.45)\times10^{-5}$\\
B [J$^{\frac{1}{4}}$·m] & $(3.81 \pm 0.34)\times10^{-15}$ & $(3.11 \pm 0.25)\times10^{-15}$ & $(2.39 \pm 0.08)\times10^{-15}$ \\
C [K$^{\frac{1}{4}}$] & $76.36 \pm 0.71$ & $77.37 \pm 0.60$ & $78.17 \pm 0.45$ \\
T$_{0} =$ C$^{4}$ [K] & $(3.40 \pm 0.13)\times10^{7}$ & $(3.58 \pm 0.11)\times10^{7}$ & $(3.73 \pm 0.09)\times10^{7}$ \\
$\chi^{2}$/dof & 0.263 & 1.268 & 0.713  \\
\hline
\multicolumn{4}{c}{\textbf{Poole--Frenkel Emission}} \\
\hline
Parameter & 200~K & 165~K & 93~K \\
\hline
\noalign{\vskip 1pt}
Slope = $\beta_{PF}/k_{B}T$ [m$^{1/2}$V$^{-1/2}$]& $(1.73 \pm 0.01)\times10^{-4}$ & $(2.28 \pm 0.93)\times10^{-4}$ & $(8.07 \pm 0.01)\times10^{-4}$\\
Intercept = ln($A\sigma_{PF}$) & $-48.23 \pm 0.71$ & $-49.52 \pm 0.61$ & $-57.16 \pm 1.28$\\
$\beta_{\mathrm{PF}}$ [$\mathrm{J \cdot m^{1/2} \, V^{-1/2}}$] & $(4.79 \pm 0.30)\times10^{-25}$ & $(5.19 \pm 0.21)\times10^{-25}$ & $(1.04 \pm 0.18)\times10^{-24}$\\
$\varepsilon_{r}$ from $\beta_{\mathrm{PF}}$ & $644.6 \pm 80.3$ & $549.2 \pm 44.6$ & $137.6 \pm 48.5$ \\
$R^{2}$ & 0.996 & 0.990 & 0.990 \\
\hline
\multicolumn{4}{c}{\textbf{Thermally-Assisted Tunneling}} \\

\hline
Parameter & 200~K & 165~K & 93~K \\
\hline
\noalign{\vskip 2pt}

Slope = q·a/$k_{B}T$ [m V$^{-1}$] & $(1.262 \pm 0.013)\times10^{-8}$ & $(1.55 \pm 0.13)\times10^{-8}$ & $(4.13 \pm 0.14)\times10^{-8}$\\
\noalign{\vskip 1pt}

Intercept = ln($\sigma_{0}A$) & $-33.831 \pm 0.006$ & $-34.89 \pm 0.05$ & $-39.44 \pm 0.11$\\

$a_{T}$ from slope [nm] & $0.218 \pm 0.002$ & $0.220 \pm 0.018$ & $0.331 \pm 0.011$\\

$R^{2}$ & 0.9997 & 0.9800 & 0.9907 \\

\hline
\multicolumn{4}{c}{\textbf{Phonon-Assisted Tunneling}} \\
\hline
Parameter & 200~K & 165~K & 93~K \\
\hline
\noalign{\vskip 1pt}
Slope = $q^{2}\tau^{3}/3m^{*}\hbar$ [m$^{2}$V$^{-2}$] & $(1.68 \pm 0.40)\times10^{-16}$ & $(2.01 \pm 0.42)\times10^{-16}$ & $(2.78 \pm 0.30)\times10^{-16}$\\
Intercept = ln(I$_{0}A$) & $-47.45 \pm 0.10$ & $-48.46 \pm 0.11$ & $-51.79 \pm 0.22$\\
$\tau$ from slope ($m^{*} = m_{e}$) [fs] & 12.1  & 12.9  &  14.5  \\
$R^{2}$ & 0.978 & 0.919 & 0.970 \\
\hline
\multicolumn{4}{c}{\textbf{Fowler--Nordheim Tunneling}} \\
\hline
Parameter & 200~K & 165~K & 93~K \\
\hline
\noalign{\vskip 1pt}
Slope = $-8\pi\sqrt{2m^{*}}\varphi^{3/2}/3q h$ [Vm$^{-1}$] & -- & -- & $(-1.50 \pm 0.43)\times10^{8}$\\
Intercept = ln(I$_{0}A$) & -- & -- & $-66.21 \pm 0.56$ \\
$R^{2}$ & -- & -- & 0.980\\
\hline
\hline
\end{tabular}
\caption{Fitted parameters for field-enhanced transport and tunneling models at 200~K, 165~K and 93~K}
\label{tab:summary_params}
\end{table*}

\section{Discussion}
\label{sec:discussion}

A limitation of the present data set is that the PI blocking layer covers the IDE region, influencing signal formation and interfacial charge relaxation in ways that are not independently constrained by these measurements. In the lateral devices, the PI acted as both a hole- and electron-blocking boundary, and any hole-blocking at the collecting side may reduce the measured response. Leakage through the PI and charge relaxation at the PI--a-Se interface may vary with temperature, changing the overall response amplitude. Interfacial charge buildup may also alter the local field.

The Onsager fits are therefore used as an empirical test of the expected field dependence in the FAD interval, and the fitted scale factor is interpreted as an effective amplitude parameter for the device response over the fitted range rather than as a measurement of intrinsic photogeneration alone. The EQE field dependence in the FAD interval was well reproduced by an Onsager dissociation fit, with the temperature-specific scale factor \(\xi_{\mathrm{eff}}(T)\) accounting for the overall EQE amplitude over the fitted interval. Likewise, the fitted \(r_{0}\) should be interpreted as an effective post-thermalization electron--hole separation that sets the dissociation probability. In this model, a larger \(r_{0}\) corresponds to weaker Coulomb binding and therefore a higher escape probability. This distinction helps explain why \(r_{0}(297\,\mathrm{K})=3.07\pm0.14\,\mathrm{nm}\) is smaller than the \(\sim7\,\mathrm{nm}\) value reported for 400\,nm excitation~\cite{Hijazi2016,Pai1975}. 

The two-parameter Onsager fits exhibit substantial parameter covariance. \(\xi_{\mathrm{eff}}\) and \(r_{0}\) are strongly anticorrelated, with correlation coefficients \(\rho(\xi_{\mathrm{eff}},r_{0})=-0.75\), \(-0.79\), \(-0.82\), and \(-0.94\) at 93, 165, 200, and 297\,K, respectively. This anticorrelation indicates that mechanisms which suppress the apparent yield, and therefore reduce \(\xi_{\mathrm{eff}}\), can be partially compensated in the fit by larger inferred \(r_{0}\). Several effects may contribute. Because 401\,nm absorption occurs within \(\sim 36\,\mathrm{nm}\) of the surface, carrier generation near the a-Se/vacuum interface may experience enhanced Coulomb binding due to electrostatic boundary effects, increasing the probability of geminate recombination. In contrast, edge and interface loss channels primarily reduce collection efficiency and therefore suppress \(\xi_{\mathrm{eff}}\). Through the coupled \(\xi_{\mathrm{eff}}\)--\(r_{0}\) fit, such losses may indirectly shift the extracted \(r_{0}\) to larger values in the fit.

The fitted \(r_{0}\) values decreased with temperature from \(3.07\,\mathrm{nm}\) at 297\,K to \(1.42\,\mathrm{nm}\) at 93\,K, as reported in Table~\ref{tab:Onsager_results}. Within the Onsager framework, the temperature dependence enters through
\begin{equation}
\gamma=\frac{q E r_{0}}{2 k_{B} T}
\qquad\text{and}\qquad
r_{c}=\frac{q^{2}}{4\pi\varepsilon_{r}\varepsilon_{0} k_{B} T},
\end{equation}
so that decreasing temperature increases \(r_{c}\) and strengthens geminate Coulomb attraction, while \(\gamma\) increases for fixed \(E\) and \(r_{0}\), reflecting a stronger role of the applied field relative to thermal energy in the dissociation probability. At lower temperature, thermal activation for hopping and detrapping is suppressed, increasing carrier localization and trap capture during the earliest post-thermalization interval. This can reduce the effective separation relevant for Onsager dissociation. These physical changes act in the same direction as the Onsager kernel temperature scaling, so the observed decrease in fitted \(r_{0}\) with temperature is consistent with stronger binding and enhanced localization in the cryogenic regime. For this reason, the Onsager analysis is restricted to the FAD interval, and the reported \(\xi_{\mathrm{eff}}(T)\) and \(r_{0}\) should be interpreted as effective fit parameters for that regime.

The \(I/E\) diagnostic tracks the transport product \(p_{\mathrm{free}}\mu_{\mathrm{eff}}\) up to the constant scale factor \(qA_{\mathrm{eff}}\), and the regimes are identified from the field dependence within each response. At 165\,K and 200\,K, \(I/E\) rises at low field and then forms a clear plateau. The rise indicates strong field enhancement of the transport product, consistent with field-assisted detrapping or an increase in the mobile carrier population. The plateau indicates that \(p_{\mathrm{free}}\mu_{\mathrm{eff}}\) becomes weakly field-dependent prior to multiplication, consistent with saturation of drift velocity or of the mobile carrier population. At 93\,K no plateau is observed. Instead \(I/E\) increases monotonically, consistent with strongly suppressed thermal detrapping so that the mobile carrier population remains small until field-assisted mechanisms such as hopping or phonon-assisted tunneling enhance carrier release. At 297\,K, \(I/E\) decreases across the pre-avalanche window, suggesting a reduction in \(p_{\mathrm{free}}\mu_{\mathrm{eff}}\) with increasing field, which may reflect enhanced recombination or collection limitations in the lateral geometry. This behavior is expected because as temperature decreases, thermal detrapping is strongly suppressed, so larger fields are required to field-assist emission from localized traps and populate the mobile hole reservoir. The response transitions into the AV interval when the hole carriers acquire sufficient energy for impact ionization, marked by the rapid super-exponential rise in current.

Model comparisons in the FAD interval are more physically consistent with field-assisted hopping and thermally assisted tunneling than with a Poole--Frenkel description because the extracted PF coefficient \(\beta_{\mathrm{PF}}\) disagrees with the expected value for \(\varepsilon_r=6.7\). The PF fits are formally linear in \(\ln(I/E)\) versus \(\sqrt{E}\), yet return \(\beta_{\mathrm{PF}}\) values an order of magnitude below the theoretical coefficient, which would imply unphysical permittivities if interpreted literally. This illustrates that PF linearization is not uniquely diagnostic of a PF mechanism in disordered materials, and agreement with expected parameter values provides the more informative criterion in this case. Such deviations are well known, since the standard one-dimensional PF model overestimates the barrier-lowering coefficient in disordered materials~\cite{Juhasz_1985}. For completeness, the three-dimensional Hartke formulation was also tested, but it offered no improvement and produced \(\beta_{\mathrm{PF}}\) values comparable to those from the conventional analysis~\cite{Hartke}.

Field-assisted hopping yielded characteristic \(T_0\) values in the expected range of \(10^{6}\)--\(10^{8}\,\mathrm{K}\) for amorphous semiconductors, with \(T_0 \simeq (3.4\text{--}3.7)\times10^{7}\,\mathrm{K}\) across all three temperatures. The fitted hopping parameter \(B=(\alpha\pi N_i)^{-1/4}\) is broadly consistent with an estimate based on literature values of \(N_i\) and \(a\), differing by \((44.4\text{--}65.1)\%\). This level of agreement, together with the stability of \(T_0\) across temperature, supports field-assisted hopping as a self-consistent description of the FAD-region transport.

The TAT fits were strongly linear at all three cryogenic temperatures and yielded sub-nanometer effective tunneling distances that increase at the lowest temperature, consistent with reduced thermal assistance requiring tunneling through a larger fraction of the barrier at lower temperature. The extracted \(a_T\) values are smaller than the \(1.2\,\mathrm{nm}\) reported for vertically structured a-Se devices, by roughly a factor of \(3.6\)--\(5.5\) over the measured temperatures~\cite{Kabir2014}.

The PAT fits provided acceptable linearity and yielded tunneling times of order \(10\,\mathrm{fs}\) with limited temperature dependence. Varying the assumed effective mass from \(0.1\,m_e\) to \(2.0\,m_e\) shifts the inferred \(\tau\) only modestly, although the PAT fit quality is poorer than the hopping and TAT fits.

The FN analysis did not support FN tunneling as a dominant pre-avalanche mechanism. Only the 93\,K data were compatible with the FN model, and the inferred barrier height \(\varphi \approx 74\,\mathrm{meV}\) is only \(\sim 9\,k_{B}T\) at 93\,K, implying that thermal emission across the barrier would not be strongly suppressed. This is inconsistent with interpreting the field dependence as FN tunneling through a bulk triangular barrier.

As discussed in Sec.~\ref{sec:AVLD}, lateral devices measure the gap-averaged multiplication factor \(\langle M\rangle\) given by Eq.~\ref{eq:Mavg}, which can reduce the measured AV multiplication relative to vertical devices in which carriers traverse a fixed thickness. Avalanche multiplication in a-Se was described well by the LD model, which provides a consistent basis for comparing IICs across temperatures and device geometries. For comparison, Tsuji's benchmark IIC data were measured using a vertical device with Au and Al electrodes and a CeO$_2$ hole-blocking layer, with a-Se thicknesses of 0.5 and 1.0\,\textmu m~\cite{Tsuji1989}. Although the Tsuji series stops at 148\,K, it remains the natural reference because it spans multiple temperatures under similar optical excitation and enables a direct consistency check against Kasap's LD analysis at room temperature~\cite{Kasap2004}. Two LD parameterizations are used. The energy-independent low-field form provides a compact baseline but can force the fitted energy-relaxation length $\lambda_E$ into values that are not physically plausible. This motivates the energy-dependent relaxation-length form used by Kasap, which yields physically plausible $\lambda_E(\mathcal{E}_I)$ values.

Across the overlapping temperatures, both LD analyses indicate that impact ionization is suppressed as temperature decreases in the Tsuji benchmark and in the IIC data from this work. In the energy-independent low-field LD form, $\mathcal{E}_I$ and $\lambda_E$ are strongly correlated and are therefore only weakly identifiable for sparse data sets, so fixing one parameter forces the observed temperature dependence of the IICs to be absorbed by the other. When $\lambda_E$ is held fixed, the suppression is captured as an increase in the fitted effective threshold $\mathcal{E}_I$ with decreasing temperature. When $\mathcal{E}_I$ is instead held fixed at $0.66\,\mathrm{eV}$, the same suppression is captured as a systematic decrease in the fitted $\lambda_E$ with decreasing temperature. The energy-dependent LD analysis addresses a different assumption by allowing the energy-relaxation length to vary with carrier energy through $\lambda_E(\mathcal{E})=\lambda_{E0}+\beta\,\mathcal{E}^{n}$, and it yields physically plausible $\lambda_E(\mathcal{E}_I)$ values while preserving the same qualitative temperature dependence inferred from the energy-independent form.

A clear deviation from the otherwise smooth behavior appeared in the intermediate-temperature window of the IIC data from this work. The 165\,K AV growth is anomalously slow, and the limited 200\,K points show similarly slow growth in Fig.~\ref{fig:LDPlot} (top), indicating a suppression of net multiplication in the 165--200\,K range. Relative to this intermediate-temperature suppression, the 93\,K response is consistent with a recovery in which net multiplication increases once the suppressing mechanism becomes less effective. This behavior is evident in the gain and IIC curves in Fig.~\ref{fig:LDPlot} and is reproducible across repeated measurements and samples. The anomaly also appears in both the energy-independent and energy-dependent LD analyses, indicating that it reflects a real reduction of net multiplication in this temperature range rather than a fit artifact tied to a single LD parameterization.

A plausible explanation is a crossover in the pre-avalanche transport regime that changes the carrier energy distribution entering multiplication. Although trapping generally strengthens as temperature decreases, the relevant quantity for AV is the extent to which carriers experience repeated energy-resetting interruptions during high-field transport. In the 165--200\,K window, a transport pathway that increases the frequency or effectiveness of such interruptions would reduce the high-energy tail and slow the rise of multiplication with field. At 93\,K, the disappearance of the TL branch suggests that conduction shifts toward a more field-driven process that shortens the effective trapping dwell relative to the transit time or allows carriers to regain energy more efficiently between inelastic loss events, thereby restoring a larger high-energy carrier fraction and increasing the net ionization probability relative to the intermediate-temperature case.

An alternative, non-exclusive possibility is that the inferred suppression at intermediate temperature reflects changes in the effective field profile rather than changes in the intrinsic IIC. Temperature-dependent space charge, polarization, or field localization can alter the local peak field in the multiplication zone while leaving the nominal applied field unchanged. Such field redistribution would bias the extracted IIC when evaluated under a uniform-field assumption, and it naturally links changes in the pre-avalanche response to the observed suppression of multiplication in the 165--200\,K window and its apparent recovery at 93\,K.

\begin{figure}
\centering
\includegraphics[width=0.95\columnwidth]{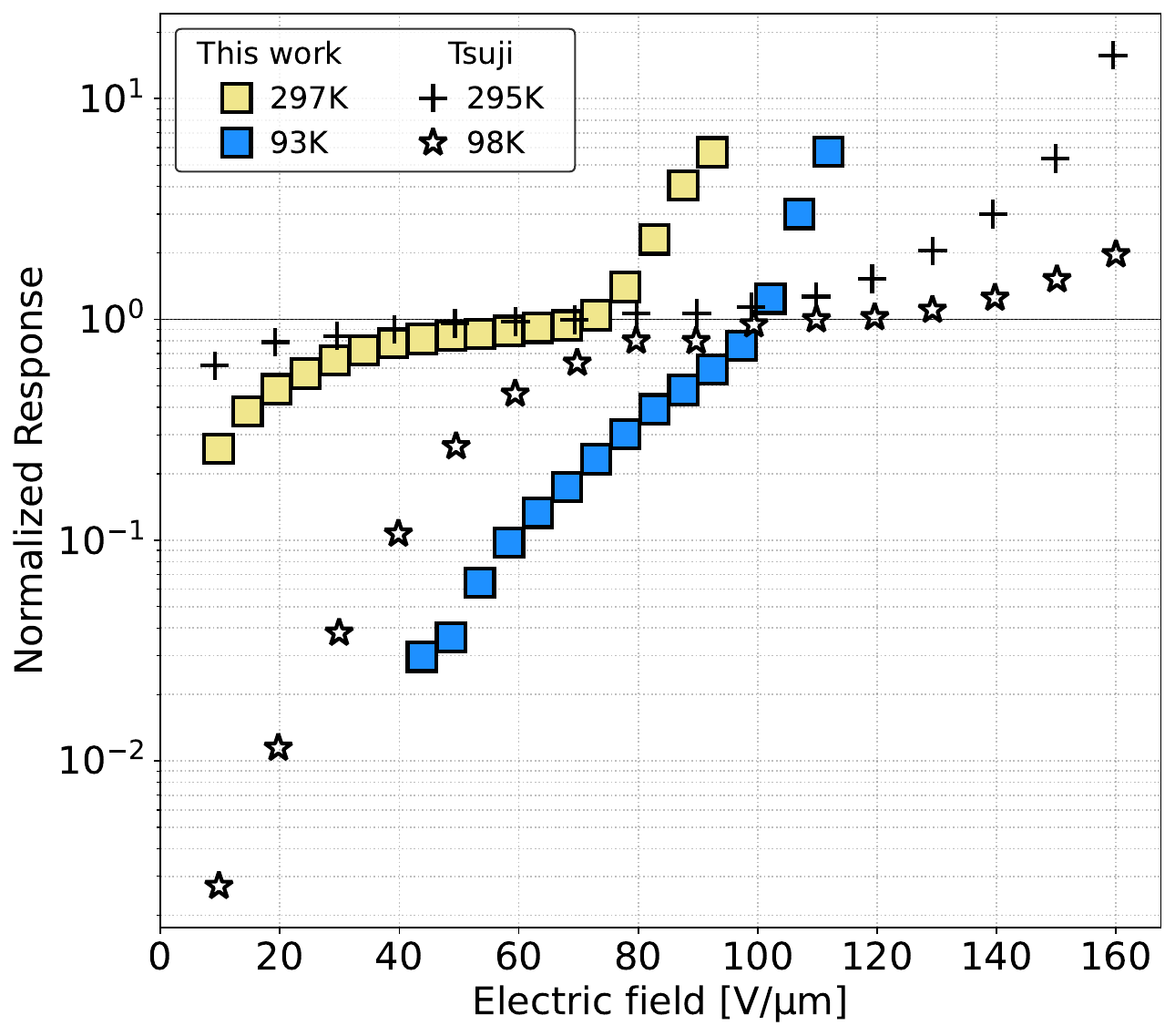}
\caption{Normalized response for lateral a-Se devices from this work at 297~K and 93~K and for vertical devices at 295~K and 98~K extracted from Tsuji~\cite{Tsuji1994}. Each curve is normalized at its avalanche-onset field such that the response equals unity at onset and values above unity indicate hole multiplication.}
\label{fig:response1}
\end{figure}

Tsuji later reported several temperature-dependent hole-response curves down to 98\,K under 400\,nm excitation using the same vertical device architecture~\cite{Tsuji1994}. This enables a qualitative comparison to the 93\,K lateral-device response. In Fig.~\ref{fig:response1}, the room-temperature and cryogenic responses from this work and from Tsuji are normalized at the avalanche-onset field such that values above unity correspond to hole multiplication. Two features stand out. First, the avalanche-onset fields are closely aligned between the vertical and lateral devices at both temperatures: the onsets at 295\,K and 297\,K agree well, and the cryogenic onsets at 98\,K and 93\,K also agree, making the temperature-driven shift in onset field clear in both geometries. Second, Tsuji's cryogenic curve exhibits a TL region below onset, whereas the 93\,K lateral response does not. If deep-cryogenic transport in the lateral geometry becomes more field-driven and less frequently interrupted by energy-resetting loss events, multiplication can become measurable immediately upon turn-on, collapsing the field interval in which collection would otherwise saturate into a distinct TL plateau. In contrast, the short drift distance in the thin vertical device can allow collection to saturate before multiplication becomes appreciable, preserving a pre-onset TL branch.

Limitations arise from the lateral geometry, the single-wavelength excitation, and the use of one device per temperature, which can introduce device-to-device variation in the absolute scale of the response. In addition, \(\xi_{\mathrm{eff}}\) is an effective scale factor that absorbs collection losses over the fitted interval, so direct extraction of \(\eta\) requires independent collection measurements or time-resolved studies, and space charge and injection at the highest fields complicate quantitative inference in the screened regime. Despite these caveats, the combined use of the \(I/E\) diagnostic, mechanism-specific linearizations, and the LD model yields a coherent interpretation across temperature.

Several implications for low-temperature photon detection follow. First, the regime boundaries shift with temperature. The FAD$\rightarrow$TL transition and the onset of AV move to higher applied fields as temperature decreases, and no clear TL plateau is observed at 93\,K, as shown in Fig.~\ref{fig:QPlot}. Second, the LD analysis indicates that energy relaxation remains efficient on nanometer length scales in the avalanche regime, and the extracted temperature dependences of the effective threshold and energy-relaxation parameters are consistent with progressively suppressed impact ionization at lower temperature. Third, operation should avoid the high-field screened regime to preserve stable gain.

\section{Conclusions}
\label{sec:conclusions}

This work reports the field-dependent photoresponse of laterally structured a-Se devices from 297\,K to 93\,K under 401\,nm excitation. The pre-avalanche response exhibited distinct field regimes whose boundaries shifted to higher applied fields as temperature decreases. A TL plateau was evident at 200\,K and 165\,K, but it was absent at 93\,K where the response transitions directly from FAD into AV.

Over the FAD interval, the measured EQE field dependence was reproduced by fitting the Onsager dissociation model using the effective parameters \(\xi_{\mathrm{eff}}(T)\) and \(r_{0}\). The fitted \(r_{0}\) decreased with decreasing temperature, and the strong anticorrelation between \(\xi_{\mathrm{eff}}\) and \(r_{0}\) indicates that the overall EQE scale and dissociation are not separable in the present measurement. In the same interval, pre-avalanche transport is most consistent with field-assisted hopping and thermally assisted tunneling based on agreement with expected parameter values, while Poole--Frenkel and Fowler--Nordheim interpretations are disfavored.

Avalanche multiplication was described by the LD model across temperatures. Relative to room temperature, the extracted impact-ionization coefficients decreased as temperature was reduced. Within the cryogenic data from this work, multiplication growth was suppressed in the 165--200\,K interval and increased again at 93\,K. For low-temperature photon-detection operation, these results indicate that the fields required to reach controlled gain increase as temperature decreases, and stable gain requires avoiding the screened regime.

\begin{acknowledgments}
This work was supported by the U.S. Department of Energy through the Office of Science Graduate Student Research (SCGSR) program under contract DE-SC0014664. The author gratefully acknowledges the support and resources provided by the Physics Division at Oak Ridge National Laboratory. This material is also based upon work supported by the U.S. Department of Energy, Office of Science, Office of High Energy Physics under Award Number DE-SC0020065. Parts of this research were funded by the National Science Foundation under award number NSF-PHY2412808 (FSU) and STFC grant number ST/W003945/1.
\end{acknowledgments}

\bibliographystyle{unsrtnat}
\bibliography{Bibliography}

@article{kasap2009,
author = {Kasap, Safa and Frey, Joel B. and Belev, George and Tousignant, Olivier and Mani, Habib and Laperriere, Luc and Reznik, Alla and Rowlands, John A.},
title = {Amorphous selenium and its alloys from early xeroradiography to high resolution X-ray image detectors and ultrasensitive imaging tubes},
journal = {physica status solidi (b)},
volume = {246},
number = {8},
pages = {1794-1805},
doi = {https://doi.org/10.1002/pssb.200982007},
url = {https://onlinelibrary.wiley.com/doi/abs/10.1002/pssb.200982007},
eprint = {https://onlinelibrary.wiley.com/doi/pdf/10.1002/pssb.200982007},
year = {2009}
}

@article{McKenzie1987,
doi = {10.1088/0268-1242/2/5/005},
url = {https://doi.org/10.1088/0268-1242/2/5/005},
year = {1987},
month = {may},
publisher = {},
volume = {2},
number = {5},
pages = {275},
author = {S McKenzie and M G Burt},
title = {A test by Monte Carlo simulation of the lucky drift theory of impact ionisation for a model with energy-dependent parameters},
journal = {Semiconductor Science and Technology}
}

@article{Ridley1983,
doi = {10.1088/0022-3719/16/17/020},
url = {https://doi.org/10.1088/0022-3719/16/17/020},
year = {1983},
month = {jun},
publisher = {},
volume = {16},
number = {17},
pages = {3373},
author = {B K Ridley},
title = {Lucky-drift mechanism for impact ionisation in semiconductors},
journal = {Journal of Physics C: Solid State Physics}
}

@article{Shockley1961,
  author  = {Shockley, William},
  title   = {Problems Related to $p$--$n$ Junctions in Silicon},
  journal = {Solid-State Electronics},
  volume  = {2},
  number  = {1},
  pages   = {35--67},
  year    = {1961},
  issn    = {0038-1101},
  doi     = {10.1016/0038-1101(61)90054-5},
}

@article{Tsuji1994,
  author  = {Tsuji, Kazutaka and Takasaki, Yukio and Hirai, Tadaaki and Yamazaki, Junichi and Tanioka, Kenkichi},
  title   = {Avalanche Phenomenon in Amorphous Selenium},
  journal = {Optoelectronics---Devices and Technologies},
  volume  = {9},
  number  = {3},
  pages   = {367--378},
  year    = {1994},
  month   = {September}
}

@INPROCEEDINGS{Hijazi2014, author={Hijazi, Nour and Kabir, M. Z.}, booktitle={2014 IEEE Nuclear Science Symposium and Medical Imaging Conference (NSS/MIC)}, title={Modeling of temperature and field dependent effective hole mobility at high fields in amorphous selenium}, year={2014}, volume={}, number={}, pages={1-3}, doi={10.1109/NSSMIC.2014.7431180}}

@article{Juhasz_1985,
doi = {10.1088/0022-3727/18/4/016},
url = {https://doi.org/10.1088/0022-3727/18/4/016},
year = {1985},
month = {apr},
publisher = {},
volume = {18},
number = {4},
pages = {721},
author = {C Juhasz and S O Kasap},
title = {Charge transport in amorphous xerographic photoreceptor films of chlorine-doped Se0.995As0.005},
journal = {Journal of Physics D: Applied Physics}
}

@article{Frey,
    author = {Frey, Joel B. and Belev, George and Tousignant, Olivier and Mani, Habib and Laperriere, Luc and Kasap, Safa O.},
    title = {Dark current in multilayer stabilized amorphous selenium based photoconductive x-ray detectors},
    journal = {Journal of Applied Physics},
    volume = {112},
    number = {1},
    pages = {014502},
    year = {2012},
    month = {07},
    issn = {0021-8979},
    doi = {10.1063/1.4730135},
    url = {https://doi.org/10.1063/1.4730135},
    eprint = {https://pubs.aip.org/aip/jap/article-pdf/doi/10.1063/1.4730135/15100671/014502_1_online.pdf},
}

@article{Hartke,
    author = {Hartke, J. L.},
    title = {The Three‐Dimensional Poole‐Frenkel Effect},
    journal = {Journal of Applied Physics},
    volume = {39},
    number = {10},
    pages = {4871-4873},
    year = {1968},
    month = {09},
    issn = {0021-8979},
    doi = {10.1063/1.1655871},
    url = {https://doi.org/10.1063/1.1655871},
    eprint = {https://pubs.aip.org/aip/jap/article-pdf/39/10/4871/18347798/4871_1_online.pdf},
}

@article{juska,
author = {Juška, G. and Arlauskas, K.},
title = {Impact ionization and mobilities of charge carriers at high electric fields in amorphous selenium},
journal = {physica status solidi (a)},
volume = {59},
number = {1},
pages = {389-393},
doi = {https://doi.org/10.1002/pssa.2210590151},
url = {https://onlinelibrary.wiley.com/doi/abs/10.1002/pssa.2210590151},
eprint = {https://onlinelibrary.wiley.com/doi/pdf/10.1002/pssa.2210590151},
year = {1980}
}

@article{godet2002,
title = {Variable range hopping revisited: the case of an exponential distribution of localized states},
journal = {Journal of Non-Crystalline Solids},
volume = {299-302},
pages = {333-338},
year = {2002},
note = {19th International Conference on Amorphous and Microcrystalline Semiconductors},
issn = {0022-3093},
doi = {https://doi.org/10.1016/S0022-3093(01)01008-0},
url = {https://www.sciencedirect.com/science/article/pii/S0022309301010080},
author = {C Godet}
}

@article{Kuksenkov1998,
    author = {Kuksenkov, D. V. and Temkin, H. and Osinsky, A. and Gaska, R. and Khan, M. A.},
    title = {Origin of conductivity and low-frequency noise in reverse-biased GaN p-n junction},
    journal = {Applied Physics Letters},
    volume = {72},
    number = {11},
    pages = {1365-1367},
    year = {1998},
    month = {03},
    issn = {0003-6951},
    doi = {10.1063/1.121056},
    url = {https://doi.org/10.1063/1.121056},
    eprint = {https://pubs.aip.org/aip/apl/article-pdf/72/11/1365/18532679/1365_1_online.pdf},
}

@article{Serdouk2015,
title = {Density of states in pure and As doped amorphous selenium determined from transient photoconductivity using Laplace-transform method},
journal = {Physica B: Condensed Matter},
volume = {459},
pages = {122-128},
year = {2015},
issn = {0921-4526},
doi = {https://doi.org/10.1016/j.physb.2014.12.002},
url = {https://www.sciencedirect.com/science/article/pii/S0921452614009405},
author = {Fadila Serdouk and Mohammed Loutfi Benkhedir}
}

@article{kleinman1965,
  title = {Theory of Phonon-Assisted Tunneling in Semiconductors},
  author = {Kleinman, Leonard},
  journal = {Phys. Rev.},
  volume = {140},
  issue = {2A},
  pages = {A637--A648},
  numpages = {0},
  year = {1965},
  month = {Oct},
  publisher = {American Physical Society},
  doi = {10.1103/PhysRev.140.A637},
  url = {https://link.aps.org/doi/10.1103/PhysRev.140.A637}
}

@article{Lee2011,
    author = {Lee, Gwan-Hyoung and Yu, Young-Jun and Lee, Changgu and Dean, Cory and Shepard, Kenneth L. and Kim, Philip and Hone, James},
    title = {Electron tunneling through atomically flat and ultrathin hexagonal boron nitride},
    journal = {Applied Physics Letters},
    volume = {99},
    number = {24},
    pages = {243114},
    year = {2011},
    month = {12},
    issn = {0003-6951},
    doi = {10.1063/1.3662043},
    url = {https://doi.org/10.1063/1.3662043},
    eprint = {https://pubs.aip.org/aip/apl/article-pdf/doi/10.1063/1.3662043/14462226/243114_1_online.pdf},
}

@article{Katzenmeyer2010,
  author  = {Aaron M. Katzenmeyer and François Léonard and A. Alec Talin and Ping-Show Wong and Diana L. Huffaker},
  title   = {Poole-Frenkel Effect and Phonon-Assisted Tunneling in GaAs Nanowires},
  journal = {Nano Letters},
  volume  = {10},
  number  = {12},
  pages   = {4935--4938},
  year    = {2010},
  doi     = {10.1021/nl102958g}
}

@article{Ganichev2000,
  title = {Distinction between the Poole-Frenkel and tunneling models of electric-field-stimulated carrier emission from deep levels in semiconductors},
  author = {Ganichev, S. D. and Ziemann, E. and Prettl, W. and Yassievich, I. N. and Istratov, A. A. and Weber, E. R.},
  journal = {Phys. Rev. B},
  volume = {61},
  issue = {15},
  pages = {10361--10365},
  numpages = {0},
  year = {2000},
  month = {Apr},
  publisher = {American Physical Society},
  doi = {10.1103/PhysRevB.61.10361},
  url = {https://link.aps.org/doi/10.1103/PhysRevB.61.10361}
}

@article{tabak1968,
  title = {Field-Controlled Photogeneration and Free-Carrier Transport in Amorphous Selenium Films},
  author = {Tabak, Mark D. and Warter, Peter J.},
  journal = {Phys. Rev.},
  volume = {173},
  issue = {3},
  pages = {899--907},
  numpages = {0},
  year = {1968},
  month = {Sep},
  publisher = {American Physical Society},
  doi = {10.1103/PhysRev.173.899},
  url = {https://link.aps.org/doi/10.1103/PhysRev.173.899}
}

@article{vincent1979,
    author = {Vincent, G. and Chantre, A. and Bois, D.},
    title = {Electric field effect on the thermal emission of traps in semiconductor junctions},
    journal = {Journal of Applied Physics},
    volume = {50},
    number = {8},
    pages = {5484-5487},
    year = {1979},
    month = {08},
    issn = {0021-8979},
    doi = {10.1063/1.326601},
    url = {https://doi.org/10.1063/1.326601},
    eprint = {https://pubs.aip.org/aip/jap/article-pdf/50/8/5484/18384158/5484_1_online.pdf},
}

@article{Leiga1968,
author = {Algird G. Leiga},
journal = {J. Opt. Soc. Am.},
number = {11},
pages = {1441--1445},
publisher = {Optica Publishing Group},
title = {Optical Properties of Amorphous Selenium in the Vacuum Ultraviolet},
volume = {58},
month = {Nov},
year = {1968},
url = {https://opg.optica.org/abstract.cfm?URI=josa-58-11-1441},
doi = {10.1364/JOSA.58.001441},
}

@article{Mott1969,
author = {N. F. Mott},
title = {Conduction in non-crystalline materials},
journal = {The Philosophical Magazine: A Journal of Theoretical Experimental and Applied Physics},
volume = {19},
number = {160},
pages = {835--852},
year = {1969},
publisher = {Taylor \& Francis},
doi = {10.1080/14786436908216338},
URL = {https://doi.org/10.1080/14786436908216338},
eprint = {https://doi.org/10.1080/14786436908216338}
}

@article{Hill1971,
author = {Robert M. Hill},
title = {Hopping conduction in amorphous solids},
journal = {The Philosophical Magazine: A Journal of Theoretical Experimental and Applied Physics},
volume = {24},
number = {192},
pages = {1307--1325},
year = {1971},
publisher = {Taylor \& Francis},
doi = {10.1080/14786437108217414},
URL = {https://doi.org/10.1080/14786437108217414},
eprint = {https://doi.org/10.1080/14786437108217414}
}

@article{Pfister1977,
  author       = {Pfister, G. and Scher, H.},
  title        = {Time-dependent electrical transport in amorphous solids},
  journal      = {Physical Review B},
  year         = {1977},
  volume       = {15},
  pages        = {2062--2075},
  doi          = {10.1103/PhysRevB.15.2062}
}

@article{Nenashev2018,
  author       = {Nenashev, A. V. and Oelerich, J. O. and Jandieri, K. and Valkovskii, V. V.
                  and Semeniuk, O. and Dvurechenskii, A. V. and Gebhard, F. and Juška, G.
                  and Reznik, A. and Baranovskii, S. D.},
  title        = {Field-enhanced mobility in the multiple-trapping regime},
  journal      = {Physical Review B},
  year         = {2018},
  volume       = {98},
  pages        = {035201},
  doi          = {10.1103/PhysRevB.98.035201}
}

@article{Simmons1971,
doi = {10.1088/0022-3727/4/5/202},
url = {https://doi.org/10.1088/0022-3727/4/5/202},
year = {1971},
month = {may},
publisher = {},
volume = {4},
number = {5},
pages = {613},
author = {J G Simmons},
title = {Conduction in thin dielectric films},
journal = {Journal of Physics D: Applied Physics}
}

@article{Kabir2014,
    author = {Kabir, M. Z. and Hijazi, Nour},
    title = {Temperature and field dependent effective hole mobility and impact ionization at extremely high fields in amorphous selenium},
    journal = {Applied Physics Letters},
    volume = {104},
    number = {19},
    pages = {192103},
    year = {2014},
    month = {05},
    issn = {0003-6951},
    doi = {10.1063/1.4876239},
    url = {https://doi.org/10.1063/1.4876239},
    eprint = {https://pubs.aip.org/aip/apl/article-pdf/doi/10.1063/1.4876239/10150536/192103_1_online.pdf},
}

@article{Kasap2015,
  author    = {Safa Kasap and Cyril Koughia and Julia Berashevich and Robert Johanson and Alla Reznik},
  title     = {Charge transport in pure and stabilized amorphous selenium: re-examination of the density of states distribution in the mobility gap and the role of defects},
  journal   = {Journal of Materials Science: Materials in Electronics},
  year      = {2015},
  volume    = {26},
  number    = {7},
  pages     = {4644--4658},
  doi       = {10.1007/s10854-015-3069-1},
  url       = {https://doi.org/10.1007/s10854-015-3069-1},
  issn      = {1573-482X}
}

@article{pfister1976,
  title = {Dispersive Low-Temperature Transport in $a$-Selenium},
  author = {Pfister, G.},
  journal = {Phys. Rev. Lett.},
  volume = {36},
  issue = {5},
  pages = {271--273},
  numpages = {0},
  year = {1976},
  month = {Feb},
  publisher = {American Physical Society},
  doi = {10.1103/PhysRevLett.36.271},
  url = {https://link.aps.org/doi/10.1103/PhysRevLett.36.271}
}

@article{cohen1969,
  title = {Simple Band Model for Amorphous Semiconducting Alloys},
  author = {Cohen, Morrel H. and Fritzsche, H. and Ovshinsky, S. R.},
  journal = {Phys. Rev. Lett.},
  volume = {22},
  issue = {20},
  pages = {1065--1068},
  numpages = {0},
  year = {1969},
  month = {May},
  publisher = {American Physical Society},
  doi = {10.1103/PhysRevLett.22.1065},
  url = {https://link.aps.org/doi/10.1103/PhysRevLett.22.1065}
}

@book{mott1979,
author={Mott,N. F. S. and Davis,E. A.},
year={1979;2012;},
title={Electronic processes in non-crystalline materials},
publisher={Clarendon Press},
address={New York;Oxford;},
edition={2d;2;2nd;},
isbn={9780198512882;0198512880;9780199645336;0199645337;},
language={English},
}

@article{Tsuji1989,
  title        = {Impact ionization process in amorphous selenium},
  author       = {Tsuji, K. and Takasaki, Y. and Hirai, T. and Taketoshi, K.},
  journal      = {Journal of Non-Crystalline Solids},
  volume       = {114},
  pages        = {94--96},
  year         = {1989},
  publisher    = {Elsevier},
  doi          = {10.1016/0022-3093(89)90079-3}
}

@article{Jandieri2008,
author = {Jandieri, K. and Rubel, O. and Baranovskii, S. D. and Reznik, A. and Rowlands, J. A. and Kasap, S. O.},
title = {One-dimensional lucky-drift model with scattering and movement asymmetries for impact ionization in amorphous semiconductors},
journal = {physica status solidi c},
volume = {5},
number = {3},
pages = {796-799},
doi = {https://doi.org/10.1002/pssc.200777565},
url = {https://onlinelibrary.wiley.com/doi/abs/10.1002/pssc.200777565},
eprint = {https://onlinelibrary.wiley.com/doi/pdf/10.1002/pssc.200777565},
year = {2008}
}

@article{Rubel2004,
author = {Rubel, O. and Baranovskii, S. D. and Zvyagin, I. P. and Thomas, P. and Kasap, S. O.},
title = {Lucky-drift model for avalanche multiplication in amorphous semiconductors},
journal = {physica status solidi (c)},
volume = {1},
number = {5},
pages = {1186-1193},
doi = {https://doi.org/10.1002/pssc.200304319},
url = {https://onlinelibrary.wiley.com/doi/abs/10.1002/pssc.200304319},
eprint = {https://onlinelibrary.wiley.com/doi/pdf/10.1002/pssc.200304319},
year = {2004}
}

@article{Tanaka2014,
author = {Tanaka, Keiji},
year = {2014},
month = {03},
pages = {243-251},
title = {Avalanche breakdown in amorphous selenium(a-Se)and related materials: Brief review, critique, and proposal},
volume = {16},
journal = {Journal of Optoelectronics and Advanced Materials}
}

@article{Kasap2004,
    author = {Kasap, Safa and Rowlands, J. A. and Baranovskii, S. D. and Tanioka, Kenkichi},
    title = {Lucky drift impact ionization in amorphous semiconductors},
    journal = {Journal of Applied Physics},
    volume = {96},
    number = {4},
    pages = {2037-2048},
    year = {2004},
    month = {08},
    issn = {0021-8979},
    doi = {10.1063/1.1763986},
    url = {https://doi.org/10.1063/1.1763986},
    eprint = {https://pubs.aip.org/aip/jap/article-pdf/96/4/2037/18713379/2037\_1\_online.pdf},
}

@article{Onsager1938,
  title = {Initial Recombination of Ions},
  author = {Onsager, L.},
  journal = {Phys. Rev.},
  volume = {54},
  issue = {8},
  pages = {554--557},
  numpages = {0},
  year = {1938},
  month = {Oct},
  publisher = {American Physical Society},
  doi = {10.1103/PhysRev.54.554},
  url = {https://link.aps.org/doi/10.1103/PhysRev.54.554}
}

@article{Hijazi2016,
  author    = {Nour Hijazi and M. Z. Kabir},
  title     = {Mechanisms of charge photogeneration in amorphous selenium under high electric fields},
  journal   = {Journal of Materials Science: Materials in Electronics},
  volume    = {27},
  number    = {7},
  pages     = {7534--7539},
  year      = {2016},
  doi       = {10.1007/s10854-016-4733-9},
  url       = {https://doi.org/10.1007/s10854-016-4733-9},
  issn      = {1573-482X}
}

@article{Pai1975,
  title = {Onsager mechanism of photogeneration in amorphous selenium},
  author = {Pai, D. M. and Enck, R. C.},
  journal = {Phys. Rev. B},
  volume = {11},
  issue = {12},
  pages = {5163--5174},
  numpages = {0},
  year = {1975},
  month = {Jun},
  publisher = {American Physical Society},
  doi = {10.1103/PhysRevB.11.5163},
  url = {https://link.aps.org/doi/10.1103/PhysRevB.11.5163}
}

@article{Hijazi2015_2,
author = {Hijazi, Nour and Kabir, M.Z.},
title = {Mechanisms of temperature- and field-dependent effective drift mobilities and impact ionization coefficients in amorphous selenium},
journal = {Canadian Journal of Physics},
volume = {93},
number = {11},
pages = {1407-1412},
year = {2015},
doi = {10.1139/cjp-2015-0175},

URL = {https://doi.org/10.1139/cjp-2015-0175},
eprint = {https://doi.org/10.1139/cjp-2015-0175}
}

@article{yip1981,
    author = {Yip, Kwok‐leung and Li, Leonard S. and Chen, I.},
    title = {On the computation of Onsager quantum efficiency},
    journal = {The Journal of Chemical Physics},
    volume = {74},
    number = {1},
    pages = {751-753},
    year = {1981},
    month = {01},
    issn = {0021-9606},
    doi = {10.1063/1.440791},
    url = {https://doi.org/10.1063/1.440791},
    eprint = {https://pubs.aip.org/aip/jcp/article-pdf/74/1/751/18925963/751\_1\_online.pdf},
}

@ARTICLE{Abbaszadeh2013_2,
  author={Abbaszadeh, Shiva and Allec, Nicholas and Karim, KarimS.},
  journal={IEEE Sensors Journal}, 
  title={Characterization of Low Dark-Current Lateral Amorphous-Selenium Metal-Semiconductor-Metal Photodetectors}, 
  year={2013},
  volume={13},
  number={5},
  pages={1452-1458},
  doi={10.1109/JSEN.2012.2234450}}

@ARTICLE{Wang2010,
  author={Wang, Kai and Chen, Feng and Allec, Nicholas and Karim, Karim S.},
  journal={IEEE Transactions on Electron Devices}, 
  title={Fast Lateral Amorphous-Selenium Metal–Semiconductor–Metal Photodetector With High Blue-to-Ultraviolet Responsivity}, 
  year={2010},
  volume={57},
  number={8},
  pages={1953-1958},
  doi={10.1109/TED.2010.2051370}}

@article{Chang2016,
    author = {Chang, Cheng-Yi and Pan, Fu-Ming and Lin, Jian-Siang and Yu, Tung-Yuan and Li, Yi-Ming and Chen, Chieh-Yang},
    title = {Lateral amorphous selenium metal-insulator-semiconductor-insulator-metal photodetectors using ultrathin dielectric blocking layers for dark current suppression},
    journal = {Journal of Applied Physics},
    volume = {120},
    number = {23},
    pages = {234501},
    year = {2016},
    month = {12},
    issn = {0021-8979},
    doi = {10.1063/1.4972029},
    url = {https://doi.org/10.1063/1.4972029},
    eprint = {https://pubs.aip.org/aip/jap/article-pdf/doi/10.1063/1.4972029/15187587/234501\_1\_online.pdf},
}

@article{Rooks2026,
doi = {10.1088/1748-0221/21/01/P01032},
url = {https://doi.org/10.1088/1748-0221/21/01/P01032},
year = {2026},
month = {jan},
publisher = {IOP Publishing},
volume = {21},
number = {01},
pages = {P01032},
author = {Rooks, M. and Abbaszadeh, S. and Asaadi, J. and Chirayath, V.A. and Febbraro, M. and García-Peris, M.Á. and Gramellini, E. and Hellier, K. and Sudarsan, B. and Tzoka, I.},
title = {Characterization of lateral amorphous selenium photodetectors for low-photon and VUV detection at cryogenic temperatures},
journal = {Journal of Instrumentation}
}

@article{Rooks2023,
doi = {10.1088/1748-0221/18/01/P01029},
url = {https://dx.doi.org/10.1088/1748-0221/18/01/P01029},
year = {2023},
month = {jan},
publisher = {IOP Publishing},
volume = {18},
number = {01},
pages = {P01029},
author = {Rooks, M. and Abbaszadeh, S. and Asaadi, J. and Febbraro, M. and Gladen, R.W. and Gramellini, E. and Hellier, K. and Blaszczyk, F. Maria and McDonald, A.D.},
title = {Development of a novel, windowless, amorphous selenium based photodetector for use in liquid noble detectors},
journal = {Journal of Instrumentation}
}

@article{Hellier2023,
  author    = {Kaitlin Hellier and Derek A. Stewart and John Read and Roy Sfadia and Shiva Abbaszadeh},
  title     = {Tuning Amorphous Selenium Composition with Tellurium to Improve Quantum Efficiency at Long Wavelengths and High Applied Fields},
  journal   = {ACS Applied Electronic Materials},
  year      = {2023},
  volume    = {5},
  number    = {5},
  pages     = {2678--2685},
  doi       = {10.1021/acsaelm.3c00150},
  url       = {https://doi.org/10.1021/acsaelm.3c00150},
  publisher = {American Chemical Society}
}

@article{sensors2013,
  author = {Tomoaki Masuzawa and Ichitaro Saito and Takatoshi Yamada et al.},
  title = {Development of an Amorphous Selenium-Based Photodetector Driven by a Diamond Cold Cathode},
  journal = {Sensors},
  volume = {13},
  pages = {13744--13778},
  year = {2013},
  doi = {10.3390/s131013744}
}

@article{stavro2016,
  author = {Jann Stavro and Amir H. Goldan and Wei Zhao},
  title = {SWAD: inherent photon counting performance of amorphous selenium multi-well avalanche detector},
  journal = {Proceedings of SPIE},
  volume = {9783},
  year = {2016},
  pages = {97833Q},
  doi = {10.1117/12.2217248}
}

@article{Stavro2018,
author = {Jann Stavro and Amir H. Goldan and Wei Zhao},
title = {{Photon counting performance of amorphous selenium and its dependence on detector structure}},
volume = {5},
journal = {Journal of Medical Imaging},
number = {4},
publisher = {SPIE},
pages = {043502},
year = {2018},
doi = {10.1117/1.JMI.5.4.043502},
URL = {https://doi.org/10.1117/1.JMI.5.4.043502}
}

@article{Kuvvetli2010,
title = {CZT drift strip detectors for high energy astrophysics},
journal = {Nuclear Instruments and Methods in Physics Research Section A: Accelerators, Spectrometers, Detectors and Associated Equipment},
volume = {624},
number = {2},
pages = {486-491},
year = {2010},
note = {New Developments in Radiation Detectors},
issn = {0168-9002},
doi = {https://doi.org/10.1016/j.nima.2010.03.172},
url = {https://www.sciencedirect.com/science/article/pii/S0168900210008247},
author = {I. Kuvvetli and C. Budtz-Jørgensen and E. Caroli and N. Auricchio}
}

@article{Bornefalk2010,
  title     = {Photon-counting spectral computed tomography using silicon strip detectors: a feasibility study},
  author    = {Bornefalk, Hans and Danielsson, Mats},
  journal   = {Phys. Med. Biol.},
  volume    = {55},
  number    = {7},
  pages     = {1999--2022},
  year      = {2010},
  month     = apr,
  doi       = {10.1088/0031-9155/55/7/014}
}

@article{Schlomka2008,
doi = {10.1088/0031-9155/53/15/002},
url = {https://dx.doi.org/10.1088/0031-9155/53/15/002},
year = {2008},
month = {jul},
publisher = {},
volume = {53},
number = {15},
pages = {4031},
author = {Schlomka, J P and Roessl, E and Dorscheid, R and Dill, S and Martens, G and Istel, T and Bäumer, C and Herrmann, C and Steadman, R and Zeitler, G and Livne, A and Proksa, R},
title = {Experimental feasibility of multi-energy photon-counting K-edge imaging in pre-clinical computed tomography},
journal = {Physics in Medicine \& Biology}
}

@article{Ota2021,
author={Ota, Ryosuke},
title={Photon counting detectors and their applications ranging from particle physics experiments to environmental radiation monitoring and medical imaging},
journal={Radiological Physics and Technology},
year={2021},
month={Jun},
day={01},
volume={14},
number={2},
pages={134-148},
issn={1865-0341},
doi={10.1007/s12194-021-00615-5},
url={https://doi.org/10.1007/s12194-021-00615-5}
}

\end{document}